\documentclass[12pt]{article}%
\pdfoutput=1
\usepackage[nosort]{cite}
\usepackage{graphicx}
\usepackage{multicol}
\usepackage{amsfonts}
\usepackage{amssymb}
\usepackage{amsmath}
\usepackage{singleauthor}
\usepackage{afterpage}
\usepackage{setspace}
\usepackage{verbatim}
\usepackage{color}
\usepackage{longtable}
\usepackage{float}
\usepackage{subcaption}
\usepackage{epsfig}
\usepackage{enumerate}
\usepackage{epstopdf}
\usepackage[enableskew, vcentermath]{youngtab}
\usepackage{adjustbox}
\newsavebox{\mysavebox}

\usepackage{tikz}
\usepackage[margin=1in]{geometry}
\usepackage{titletoc}%
\usepackage{hyperref}
\hypersetup{colorlinks,%
citecolor=black,%
filecolor=black,%
linkcolor=black,%
urlcolor=black,%
pdftex}
\providecommand{\U}[1]{\protect\rule{.1in}{.1in}}
\usetikzlibrary{decorations.markings}

\numberwithin{equation}{section}

\usepackage{mathtools}
\usetikzlibrary{chains}

\tikzset{node distance=2em, ch/.style={circle,draw,on chain,inner sep=2pt},chj/.style={ch,join},every path/.style={shorten >=4pt,shorten <=4pt},line width=1pt,baseline=-1ex}

\newcommand{\ba}{\begin{eqnarray}}
\newcommand{\ea}{\end{eqnarray}}

\newcommand{\tpi}{\tilde \phi_{\rm \inf}}
\newcommand{\tp}{\tilde \phi}

\DeclareMathOperator{\bfe}{{\bf e}}

\newcommand{\be}{\begin{equation}}
\newcommand{\ee}{\end{equation}}

\tikzstyle{startstop} = [rectangle, rounded corners, minimum width=3cm, minimum height=1cm,text centered, draw=black, fill=blue!10]
\tikzstyle{startstop} = [rectangle, rounded corners, minimum width=3cm, minimum height=1cm,text centered, draw=black, fill=blue!10]

\tikzstyle{io} = [trapezium, trapezium left angle=70, trapezium right angle=110, minimum width=3cm, minimum height=1cm, text centered, draw=black, fill=blue!30]

\tikzstyle{process} = [rectangle, minimum width=3cm, minimum height=1cm, text centered, draw=black, fill=orange!30]
\tikzstyle{decision} = [diamond, minimum width=3cm, minimum height=1cm, text centered, draw=black, fill=green!30]

\tikzstyle{arrow} = [thick,->,>=stealth]

\tikzset{->-/.style={decoration={
  markings,
  mark=at position #1 with {\arrow[scale=2.4]{>}}},postaction={decorate}}}

\makeatletter \@addtoreset{equation}{section} \makeatother

\begin{document}

\date{\today}

\title{Learning to Inflate \\ {\Large A Gradient Ascent Approach to Random Inflation}}

\institution{IAS}{\centerline{School of Natural Sciences, Institute for Advanced Study, Princeton, NJ 08540, USA}}

\authors{Tom Rudelius\worksat{\IAS}\footnote{e-mail: {\tt rudelius@ias.edu}}}

\abstract{ Motivated by machine learning, we introduce a novel method for randomly generating inflationary potentials. Namely, we treat the Taylor coefficients of the potential as weights in a single-layer neural network and use gradient ascent to maximize the number of $e$-folds of inflation. Inflationary potentials ``learned" in this way develop a critical point, which is typically a local maximum but may also be an inflection point. We study the phenomenology of the models along the gradient ascent trajectory, finding substantial agreement with experiment for large-field local maximum models and small-field inflection point models. For two-field models of inflation, the potential eventually learns a genuine multi-field model in which the inflaton curves significantly during the course of its descent.
 }

\maketitle

\tableofcontents

\enlargethispage{\baselineskip}

\setcounter{tocdepth}{2}

\newpage

\section{Introduction \label{sec:INTRO}}

Cosmology and machine learning have seen great advancements over the last few decades, but the first applications of machine learning to cosmology \cite{Lucie-Smith:2018smo,Wang:2017sjw,Kamdar:2015fla,Kamdar:2015iaa,Ravanbakhsh:2017bbi,Schmelzle:2017vwd,Liu:2017dzi, Cole:2017kve} (and high-energy physics, more generally \cite{Dery:2017fap,Komiske:2017ubm,Carifio:2017bov, Komiske:2018oaa,deOliveira:2017pjk,Andreassen:2018apy,deOliveira:2015xxd,Larkoski:2017jix,Fraser:2018ieu,He:2017aed,Komiske:2016rsd,Collins:2018epr,Metodiev:2017vrx, Krefl:2017yox, Aguilar-Saavedra:2017rzt, Ruehle:2017mzq, Cole:2018emh, Bull:2018uow,Wang:2018rkk,Hashimoto:2018ftp,Cunningham:2018sdj,Erbin:2018csv,Dias:2018koa, Louppe:2016ylz, Cohen:2017exh, Datta:2017lxt, Brehmer:2018kdj, Heimel:2018mkt, Farina:2018fyg}) have emerged only in the last couple of years. As technological advances deliver experimental data with greater precision and machine learning becomes more powerful, it will be increasingly important to expand the interface 
between the two fields. In this paper, we present a simple yet intriguing application of machine learning techniques to inflationary model building.

The potential $V(\phi)$ is the crucial ingredient in a model of slow-roll inflation. Unfortunately, we do not yet know the correct potential. Although experimental constraints are strong enough to rule out many possible choices, a vast region of parameter space still remains. From a top-down perspective, string theory seems to give rise to an enormous landscape of vacua, which ostensibly allows for slow-roll inflation in some locations. Although the string landscape is far too complicated for detailed investigation at present, it may someday be amenable to a statistical approach. In particular, we might assume that the inflationary potential is drawn at random from some particular distribution and use this to develop a corresponding distribution on inflationary observables.

Of course, this simply leads to another question: what is the correct distribution? In a perfect world, our understanding of the string landscape will someday improve (perhaps aided by machine learning!) to the point that we can answer this question. In the meantime, the best we can do is make educated guesses. In \cite{Tegmark:2004qd} and later \cite{Frazer:2011tg}, the inflationary potential was modeled as a sum over Fourier modes with coefficients drawn from a Gaussian distribution. In \cite{Marsh:2013qca} (and subsequently in \cite{Pedro:2016sli, Wang:2016kzp, Freivogel:2016kxc, Dias:2017gva, Paban:2018ole}), the inflationary potential was modeled locally by randomly selecting a value for the Hessian at each point along the inflationary trajectory, then gluing these together to reconstruct the full inflationary potential. In \cite{Masoumi:2016eag,Bjorkmo:2017nzd} (see also \cite{Bachlechner:2014rqa, Easther:2016ire}) the authors assumed a Gaussian distribution for the 2-point correlation function between the potential at two different field values, then used this to randomly generate Taylor coefficients at a single point in field space. Reference \cite{Masoumi:2017gmh} similarly considered a Gaussian potential by randomly selecting the potential at a discrete set of field values and smoothly interpolating between them. Each of these approaches has its advantages and disadvantages.

From the perspective of effective field theory, a natural choice for the inflationary potential is a potential of the form
\begin{equation}
V(\phi) = \Lambda^4 \sum_{i=0}^\infty  c_i \frac{\phi^i}{\Lambda^i}.
\end{equation}
In a Wilsonian framework, $\Lambda$ is set equal to the scale at which the effective theory breaks down, and one expects the coefficients $c_i$ to be $O(1)$ numbers. Expanding about a minimum at $\phi =0$ and insisting that $V(0)=0$, we can set $c_0 =c_1=0$. Constructing a random potential then amounts to randomly drawing the coefficients $c_i$, $i \geq 2$ from some distribution, which we could take to be a Gaussian with zero mean and $O(1)$ standard deviation.

Unfortunately, the scenario just described will rarely produce a viable model of inflation. The observed smallness of the inflationary power spectrum requires the potential $V(\phi)$ to be at most $O(10^{-8})$ in natural units (with the reduced Planck mass $M_{\rm Pl}$ set to 1 here and in the remainder of the paper), which means that $\Lambda$ must be at most $O(10^{-2})$. On the other hand, producing 50-60 $e$-folds of inflation at such a high energy scale requires $\phi \gtrsim O(1)$ \cite{Lyth:1996im}. For $c_i \sim O(1)$, the perturbative expansion breaks down, rendering the model invalid. Small-field models of inflation, which allow for $\phi < O(1)$, require $V(\phi)$ to be even smaller (forcing $\Lambda$ to shrink as well), and they require a significant fine-tuning of the coefficients $c_i$.

One common tactic for dealing with this problem in a relatively painless manner is to allow for distinct scales $\Lambda_v$, $\Lambda_h$ to set the height and width of the potential, respectively. Namely, we write
\begin{equation}
V(\phi) = \Lambda_v^4 \sum_{i=0}^\infty  c_i \frac{\phi^i}{\Lambda_h^i}.
\label{eq:thepotential}
\end{equation}
Such a separation of vertical and horizontal scales occurs, for instance, in models of natural inflation \cite{Freese:1990rb}. This solves the problem of the small power spectrum, as we can first select the coefficients $c_i$ and afterwards simply set $\Lambda_v$ whatever it needs to be to agree with the measured power spectrum. However, it does not solve the problem of generating sufficient inflation. In a Wilsonian framework, the scale $\Lambda_h$ cannot be larger than the scale at which new physics appears (barring additional symmetries), and we know that in our universe this cannot happen any higher than the Planck scale, at which point quantum gravity kicks in. Even in models of natural inflation, where a symmetry protects the potential against operators suppressed by $\Lambda = M_{\rm Pl}$, super-Planckian $\Lambda_h$ seems to be in tension with string theory \cite{Banks:2003sx, Rudelius:2014wla, long:2016jvd, Conlon:2016aea} and the Weak Gravity Conjecture \cite{ArkaniHamed:2006dz, Rudelius:2015xta, Montero:2015ofa, Brown:2015iha, Heidenreich:2015wga}. Generously setting $\Lambda_h =1$ and requiring $\phi < 0.5$ for perturbative control, we still require significant fine-tuning of the coefficients $c_i$ to get sufficient inflation. Indeed, \cite{Masoumi:2016eag} found that a Gaussian random potential with $\Lambda_h = 0.5$ produced more than 50 $e$-folds of inflation with a probability of around 1 in $10^5$.

One way to implement this fine-tuning is to simply fix the first few coefficients by hand and randomly draw the remainder from a Gaussian distribution so as to preserve the slow-roll conditions, as in the analysis of \cite{Bjorkmo:2017nzd}. In this paper, we consider an alternative method for tuning the coefficients, motivated by machine learning. Namely, we treat the coefficients $c_i$ as ``weights" in a single-layer neural network: starting from some randomly chosen initial values $c_i^{(0)}$, we perform a gradient ascent to maximize the ``reward function" $N_e( \{c_i \})$, the number of $e$-folds of inflation. This leads to a trajectory in weight space, each point of which corresponds to a different model of inflation. Working within the slow-roll regime, we study the phenomenology of each model along the trajectory. We further extend our analysis to two-field inflation, working within the slow-roll slow-turn approximation \cite{Peterson:2010np}.

Gradient ascent does not eliminate the fine-tuning of the inflationary potential, but it does offer a natural way to place a measure on the fine-tuned coefficients of the inflationary potential rather than artificially imposing the fine-tuning by hand. Starting from some generic (i.e. not fine-tuned) initial measure on the space of coefficients, we get a new measure after each step of the gradient ascent procedure, obtained simply by letting the initial measure flow according to the dynamical system defined by gradient ascent. By choosing our reward function to be the number of $e$-folds, we ensure that the measure eventually localizes on models that produce significant inflation. This allows us to extract predictions that can be compared with experiment.

On a philosophical level, one might say that our method of gradient ascent serves as a rudimentary model of anthropic selection effects. It has been estimated \cite{Freivogel:2005vv} that nearly 60 $e$-folds of inflation are required for suitable structure formation, so models yielding significantly less inflation may be observationally excluded on anthropic grounds. Models that produce more $e$-folds are more likely to observed than models that produce fewer $e$-folds, and from this perspective one might say that the Darwinian principle of ``survival of the fittest" prefers the former. Although our method of ``learning to inflate" by gradient ascent does not have a clear physical interpretation, it does capture the essence of the Darwinian principle.

More practically, one can view our gradient ascent approach as a way to quickly generate successful inflationary trajectories in a random landscape. This could be especially useful for generating many-field inflationary models, though for this exploratory work, we constrain ourselves to models with just one and two fields. 

In the single-field case, gradient ascent leads to a local maximum (hilltop) in around $90\%$ of trials, whereas it leads to an inflection point in the remaining $10\%$ of cases. The hilltop cases yield successful large-field models of inflation, with spectral index $n_s$ and tensor-to-scalar ratio $r$ within the $1\sigma$ and $2\sigma$ inclusion regions of \emph{Planck}. These models favor a red tilt for $n_s$ and a tensor-to-ratio larger than $0.01$. On the other hand, gradient ascent for an inflection point model is capable of generating successful models of small-field inflation. In the two-field case, gradient ascent initially produces an effectively single-field model, but eventually the inflaton path starts to curve and the the model becomes genuinely multi-field. Effectively single-field models therefore seem to correspond to saddle points of the reward function $N_e$ in the space of all two-field inflationary models, with most gradient ascent trajectories approaching the saddle point along the stable manifold before exiting along an unstable direction.

The remainder of this paper is organized as follows. In section \ref{sec:ANALYSIS}, we describe our analysis in more detail, both in the single-field case and in the two-field case. In section \ref{sec:RESULTS}, we display our results. In section \ref{sec:COMPARISON}, we compare these results with those of Gaussian Random Field (GRF) Inflation. In section \ref{sec:CONC}, we present our conclusions and speculate on possible modifications of our analysis. In a series of appendices, we review the relevant aspects of machine learning, single-field inflation, and two-field inflation that will be used throughout the remainder of the paper.



\section{Analysis \label{sec:ANALYSIS}}

In this section, we describe our analysis in more detail, first for the single-field case, then for the two-field case. The relevant aspects of gradient ascent, single-field slow-roll inflation, and two-field slow-roll inflation are reviewed in appendices.

\subsection{Single-Field Inflation}

Our starting point is the Lagrangian
\begin{equation}
\mathcal{L} = -\frac{1}{2} (\partial \phi)^2 - V(\phi),
\end{equation}
with
\begin{equation}
V(\phi) = \Lambda_v^4 \sum_{i=2}^N c_i \frac{\phi^i}{\Lambda_h^i} .
\label{eq:pot}
\end{equation}
We define $\tilde \phi = \phi/\Lambda_h$ and write,
\begin{equation}
\tilde V(\phi) =  \sum_{i=2}^N c_i  \tilde \phi^i.
\label{eq:tildeV}
\end{equation}
To maintain perturbative control, we need $\tilde \phi <1$. For our analysis, we fix the initial value to be $\tilde \phi_0 = 0.5$, and we choose $N =15$. We expect the coefficients $c_i$ to be $O(1)$ numbers, so we draw our initial values of the coefficients $c_i^{(0)}$ from a normal distribution of mean $\mu = 0$ and standard deviation $\sigma = 100$. The choice of standard deviation is meaningless provided all coefficients have the same standard deviation, since we can always absorb the scaling $\sigma \rightarrow t \sigma$ by rescaling $\Lambda_v^4 \rightarrow \Lambda_v^4/t$. Note that we need $c_2 > 0$ in order to have a minimum at $\phi=0$, and we need the potential to be monotonically increasing over the interval $(0, \tilde \phi_0)$. Thus, we throw out any initial conditions $c_k^{(0)}$ that do not satisfy these properties. Our results are qualitatively unaffected by variations of the distribution of the $c_i$ or the initial value $\tilde \phi_0$. We assume that $\Lambda_v$ is chosen to give a power spectrum of the correct amplitude, so we neglect it in our analysis.

We further define $^\sim$'ed versions of the slow-roll parameters
\begin{equation}
\tilde \epsilon_V  = \frac{1}{2} \left(  \frac{\tilde V'(\tilde\phi)}{\tilde V(\tilde\phi)}  \right)^2 ~,~ \tilde \eta_V = \frac{\tilde V''(\tilde\phi)}{\tilde V(\tilde\phi)} ~,~\tilde \xi_V = \frac{\tilde V' \tilde V'''}{\tilde V^2}~,
\label{eq:tildeslowroll}
\end{equation}
where $'$ indicates a derivative with respect to $\tilde \phi$. For an initial position $\tilde \phi_0$, the number of $e$-folds produced is then given by
\begin{equation}
N_e =  \Lambda_h^2 \int^{\tilde \phi_0}_{\tilde \phi_{\text{end}}} \frac{1}{\sqrt{2 \tilde \epsilon}} d \tilde \phi.
\label{eq:Neeq}
\end{equation}
Here, inflation ends when $\epsilon_V =  \tilde \epsilon_V / \Lambda_h^2$ becomes larger than 1 and the slow-roll approximation breaks down, but for the potential in (\ref{eq:tildeV}), this is typically very close to $\tilde \phi =0$, so to a good approximation we can integrate all the way to the origin. We can view (\ref{eq:Neeq}) either (1) as an equation for $N_e$ as a function of a fixed $\Lambda_h$, or (2) we can fix $N_e = 60$ and view it an an equation for $\Lambda_h$. In our analysis, we consider both approaches, beginning with the latter. That is, we first set $N_e =60$ and define $\Lambda_h$ by $\tilde{N}_e = 60/\Lambda_h^2$, with $\tilde{N}_e$ the integral in (\ref{eq:Neeq}). Second, we fix the value of $\Lambda_h$ (considering a range of possible values) and compute $N_e = \tilde{N}_e \Lambda_h^2$ for each one. However, we still compute phenomenology at an energy scale 60 $e$-folds before the end of inflation.

From both perspectives, the goal of our gradient ascent is to maximize the function $\tilde{N_e}$ as a function of the coefficients $c_i $ which determine the potential, so the trajectory in the space of coefficients is identical in either approach. Using our definition of $\tilde V$ in (\ref{eq:tildeV}), we have
\begin{align}
\frac{\partial}{\partial c_k} \tilde N_e &= \frac{\partial}{\partial c_k}  \int  d\tilde\phi \frac{\sum_i c_i \tilde \phi^i}{\sum_i i c_i \tilde \phi^{i-1}} \nonumber \\
&= \int d \tilde\phi \left( \frac{  \tilde \phi^k }{ \tilde V' } -   \frac{  k \tilde\phi^{k-1} }{ (\tilde V')^2 } \right).
\end{align}
With this, we perform a gradient ascent to maximize $\tilde N_e$,
\begin{equation}
c_k^{(i+1)} =c_k^{(i)}+ \eta \frac{\partial}{\partial c_k} \tilde N_e(\{ c_j^{(i)}\}),
\end{equation}
with $\eta$ the learning rate. In practice, we find that a reasonably good choice for the learning rate is $\eta = \frac{1}{100} ||\frac{\partial}{\partial c_k} \tilde N_e(\{ c_j^{(i)}\}) ||^{-3/2}$ (where $||\frac{\partial}{\partial c_k} X ||^2 :=\sum_k (\frac{\partial}{\partial c_k} X)^2$), so we adopt this value in our analysis. A constant $\eta$ would lead to problems because the gradient becomes very steep near maxima of $\tilde N_e$, and gradient ascent would frequently overshoot the maxima. This unusual situation is due to the fact that $\tilde N_e$ goes to infinity for a potential that develops a local maximum at $\tilde \phi_0$: classically, a field at a local maximum will remain there for all time. Indeed, we will see in section \ref{ssec:small} than even this choice of learning rate leads to overshooting for small-field models of inflation.

This gradient ascent leads to a trajectory in the space of the coefficients $c_i$, with each point along the trajectory corresponding to a different inflationary potential. For each such potential, we compute $\tilde \epsilon_V(\tilde \phi_0)$, $\tilde \eta_V(\tilde \phi_0)$, and $\tilde \xi_V(\tilde \phi_0)$ using (\ref{eq:tildeslowroll}). In our first approach, we then compute $\Lambda_h$ via $\Lambda_h^2 = 60/\tilde N_e$, and from this we calculate slow-roll parameters $ \epsilon_V = \tilde \epsilon_v/\Lambda_h^2 $, $\eta_V = \tilde \eta_V / \Lambda_h^2 $, $\xi_V = \tilde \xi_V / \Lambda_h^4$. In our second approach, we fix $\Lambda_h$ to a range of values and compute the slow-roll parameters using the same formulae. We stop our gradient ascent when the potential no longer increases monotonically between $\tilde \phi=0$ and $\tilde \phi =\tilde \phi_0$.


\subsection{Two-Field Inflation}

Our two-field analysis is similar to the single-field case. The Lagrangian is now
\begin{equation}
\mathcal{L} = - \frac{1}{2} \left( (\partial \phi^1)^2+(\partial \phi^2)^2 \right) - V(\phi^1,\phi^2).
\end{equation}
Here, we have 
\begin{equation}
V( \phi^i) = \Lambda_v^4 \sum_{\substack{i,j \\ 2 \leq i+j \leq N}} c_{ij} \frac{( \phi^1)^i ( \phi^2)^j }{\Lambda_h^{i+j}}
\label{eq:twofieldpot}
\end{equation}
Note that we now have a matrix of coefficients $c_{ij}$ rather than a vector $c_i$. Once again, we define $^\sim$'ed values of the fields and the potential via $\tilde \phi^i = \phi^i/\Lambda_h$,
\begin{equation}
\tilde V(\tilde \phi^i) =  \sum_{\substack{i,j \\ 2 \leq i+j \leq N}} c_{ij} ( \tilde\phi^1)^i (\tilde \phi^2)^j
\end{equation}
The biggest difference in the two-field analysis is that the path of the inflaton $\phi^i$ is now model-dependent, so we must perform both a gradient ascent in $c_{ij}$ space as well as a gradient descent in $\tilde\phi^i$ space for each choice of the $c_{ij}$. To minimize confusion, we refer to the ascent in $c_{ij}$ space as the ``trajectory," whereas we refer to the descent in $\tilde\phi^i$ as the inflaton ``path."

We fix the start of the inflaton path to $\tilde \phi_0^1 = \tilde \phi_0^2 = 0.5$. We then compute the inflaton path $\gamma$ in the slow-roll, slow-turn regime via a gradient descent. The number of $e$-folds of inflation is then given by $N_e = \Lambda_h^2 \tilde N_e$, with
\begin{align}
\tilde N_e = \Lambda_h^2 \oint_\gamma \frac{1}{\sqrt{2 \tilde \epsilon}} d \tilde \phi^i,
\end{align}
and
\begin{align}
\tilde \epsilon = \frac{\epsilon }{\Lambda_h^2}  = \frac{1}{2} \frac{|\tilde \partial_i \tilde V|^2}{\tilde V^2},
\end{align}
where $\tilde \partial_i = \frac{\partial}{\partial \tilde \phi^i}$. In the two-field case, we consider only the ``first approach," setting $N_e=60$ and viewing this as an equation for $\Lambda_h$ in terms of the parameters $c_{ij}$.

We draw our initial values for the coefficients $c_{ij}$ from a normal distribution with mean $\mu=0$ and standard deviation $\sigma = 100$. We assume that the inflaton path $\gamma$ ends at a minimum at the origin. This implies that the Hessian of the potential must be positive definite at the origin and also that the field should roll towards this minimum. If the initial $c_{ij}$ do not yield a potential that satisfies these criteria, we throw out the potential and draw new initial values for the $c_{ij}$.

For a given inflaton path, we compute the gradient
\begin{equation}
\frac{d \tilde N_e}{d c_{kl}}(\gamma) = \Lambda_h^2 \oint_\gamma \frac{d}{ d c_{kl}}\left(\frac{1}{\sqrt{2 \tilde \epsilon}} \right) d \tilde \phi^i,
\end{equation}
and from this, we compute a trajectory in $c_{ij}$ space,
\begin{equation}
c_{kl}^{(i+1)} = c_{kl}^{(i)} + \eta \frac{\partial \tilde N_e(\{c_{jm}^{(i)} \})}{\partial c_{kl}}(\gamma^{(i)}).
\end{equation}
with $\eta = 10 ||\frac{\partial}{\partial c_{kl}} \tilde N_e(\{c_{jm}^{(i)} \}) ||^{-3/2}$. Note that the path $\gamma^{(i)}$ depends on the $c_{jm}^{(i)}$, so we have to recalculate it at every step in the ascent. This, along with the fact that our potential now has significantly more terms for a fixed degree, greatly slows down computation relative to the single-field case. As a result, we take $N=10$ for the two-field case, and we note that the learning rate employed here is 1000 times larger than the one used in the single-field case. For each point in the trajectory, we compute the inflationary phenomenology using the slow-roll slow-turn approximation reviewed in appendix \ref{sec:twofieldapp}.

For future reference, we record the parameter values used in our analysis in table \ref{tab:parameters}.

\begin{table}[ptb]
\centering
\renewcommand{\arraystretch}{1.5}
\begin{tabular}
[c]{|c|c|c|c|}\hline
Variable & Parameter & Single-Field Value & Two-Field Value \\ \hline
$N$ & degree of $V(\phi)$ & 15& 10 \\ \hline
$\tilde \phi_0$ & initial value of $\tilde \phi$ & 0.5& 0.5 \\ \hline
$N_e $ & number of $e$-folds  & 60&60 \\ \hline
 $c_k^{(0)}$ & initial coefficient of $\tilde\phi^k$ & $\sim \mathcal{N}(0,\sigma^2), \sigma = 100$&$\sim \mathcal{N}(0,\sigma^2), \sigma = 100$ \\\hline
 $\eta$ & learning rate & $\frac{1}{100} ||\frac{\partial}{\partial c_k} \tilde N_e(\{ c_j^{(i)}\}) ||^{-3/2}$& $10 ||\frac{\partial}{\partial c_{jk}} \tilde N_e(\{ c_{lm }^{(i)}\}) ||^{-3/2}$ \\\hline
\end{tabular}
\renewcommand{\arraystretch}{1}
\caption{The values of the parameters used in our analysis. Minor changes to these parameters leave our results qualitatively unchanged.}%
\label{tab:parameters}%
\end{table}


\section{Results \label{sec:RESULTS}}

\subsection{Large-Field Inflation at Fixed $N_e$, Variable $\Lambda_h$}

Potentials that ``learn to inflate" by the gradient ascent procedure discussed in the previous section adopt either a local maximum or an inflection point near the initial position of the inflaton, $\tilde\phi_0$. The former occurs in $90\%$ of the trials considered, so we begin by focusing on this case. 

\begin{figure}[t!]
\begin{center}
\begin{subfigure}[b]{0.45\textwidth}
\includegraphics[trim={0cm 0cm 0cm 0cm},clip,scale=0.5]{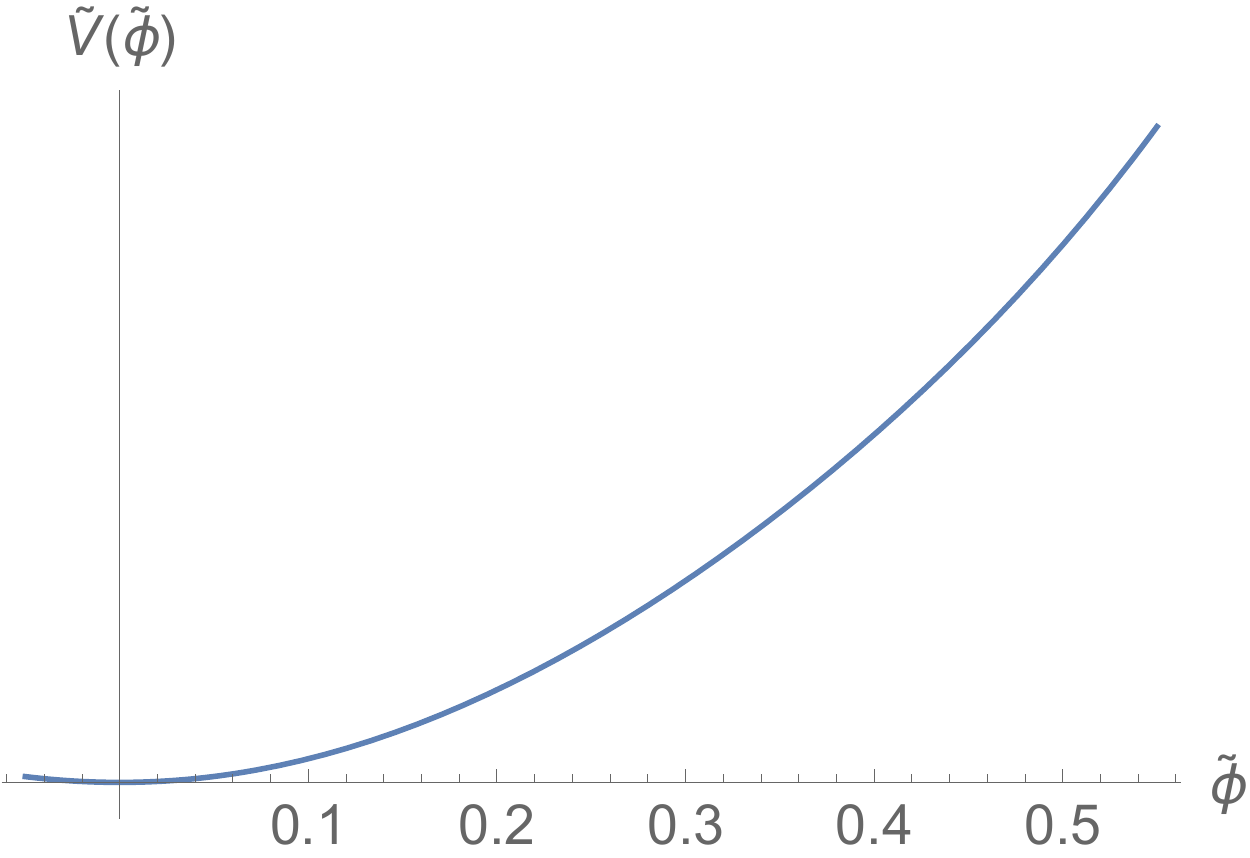}
\caption{~}
\end{subfigure}
~~~~
\begin{subfigure}[b]{0.45\textwidth}
\includegraphics[trim={0cm 0cm 0cm 0cm},clip,scale=0.5]{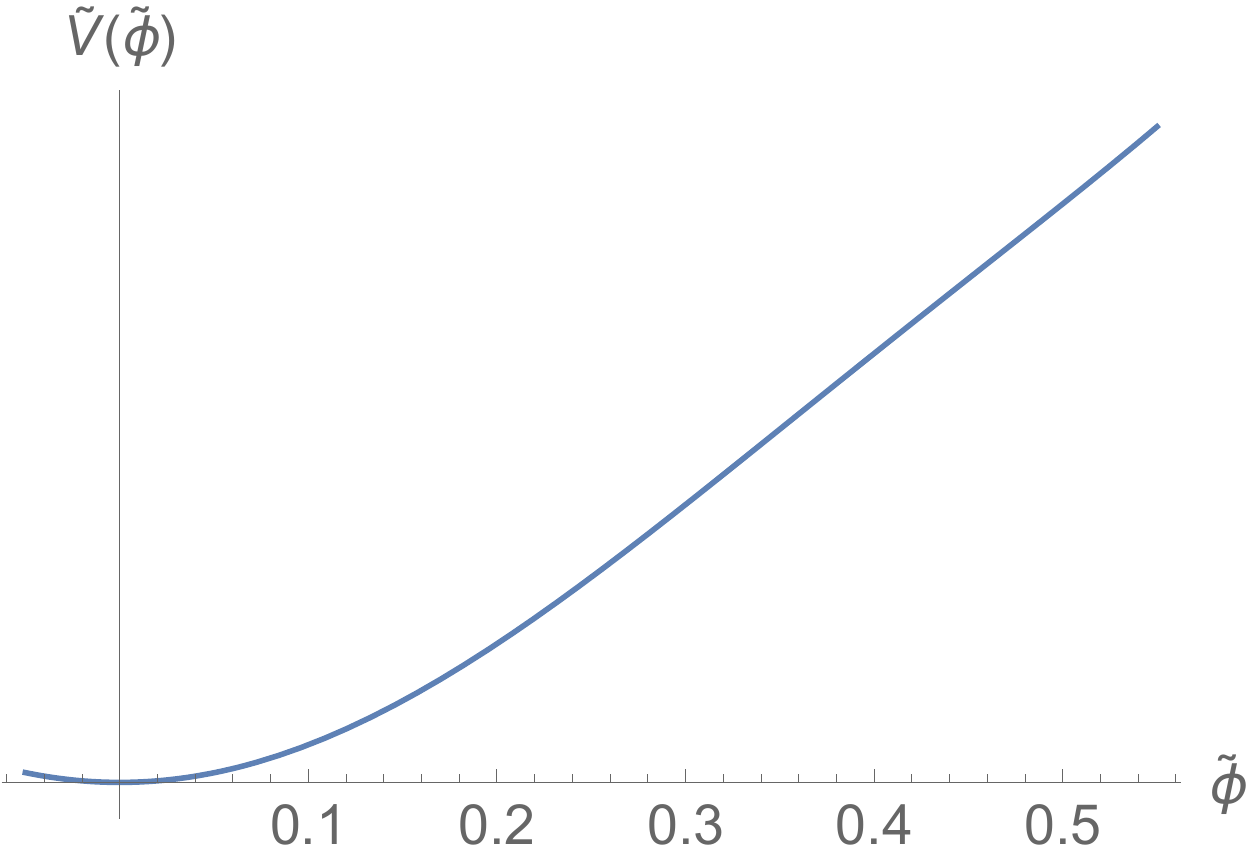}
\caption{~}
\end{subfigure}
\begin{subfigure}[b]{0.45\textwidth}
\includegraphics[trim={0cm 0cm 0cm 0cm},clip,scale=0.5]{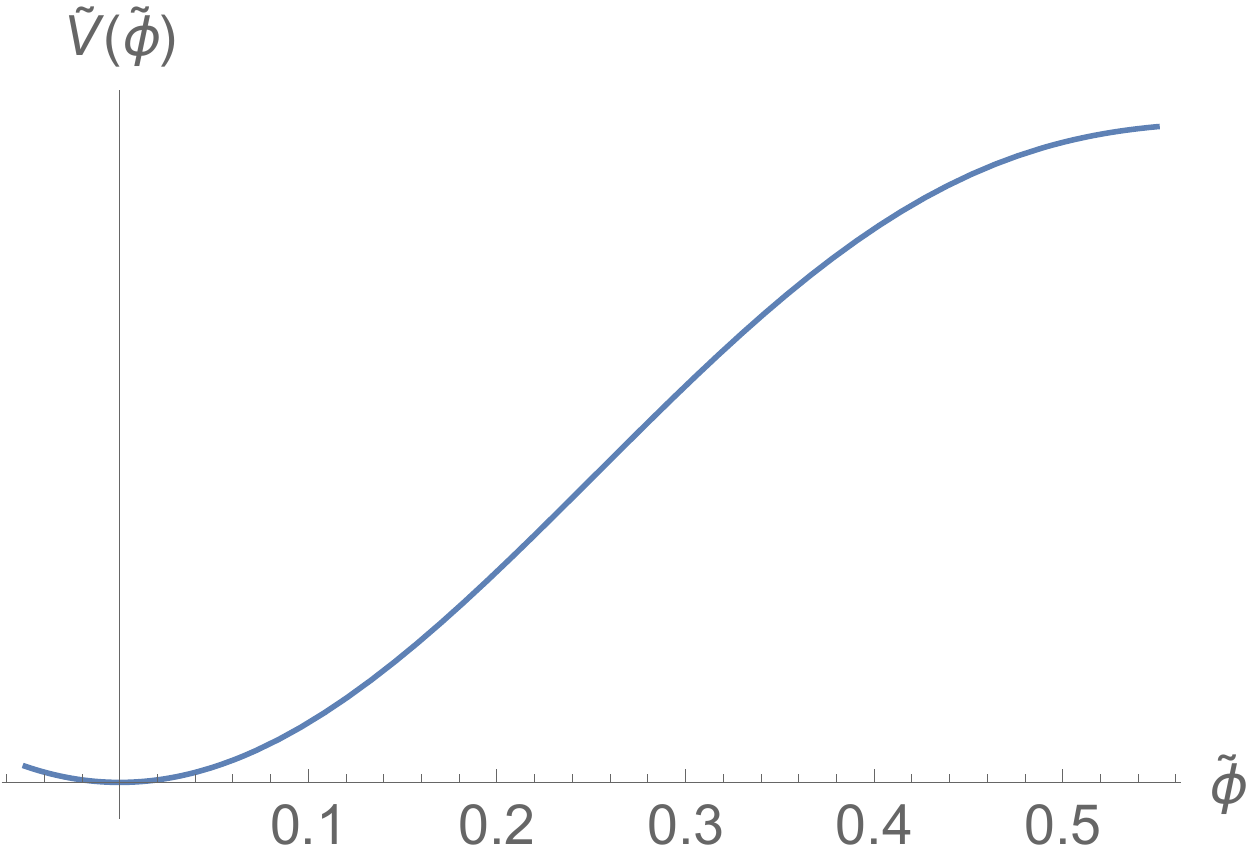}
\caption{~}
\end{subfigure}~~~~
\begin{subfigure}[b]{0.45\textwidth}
\includegraphics[trim={0cm 0cm 0cm 0cm},clip,scale=0.5]{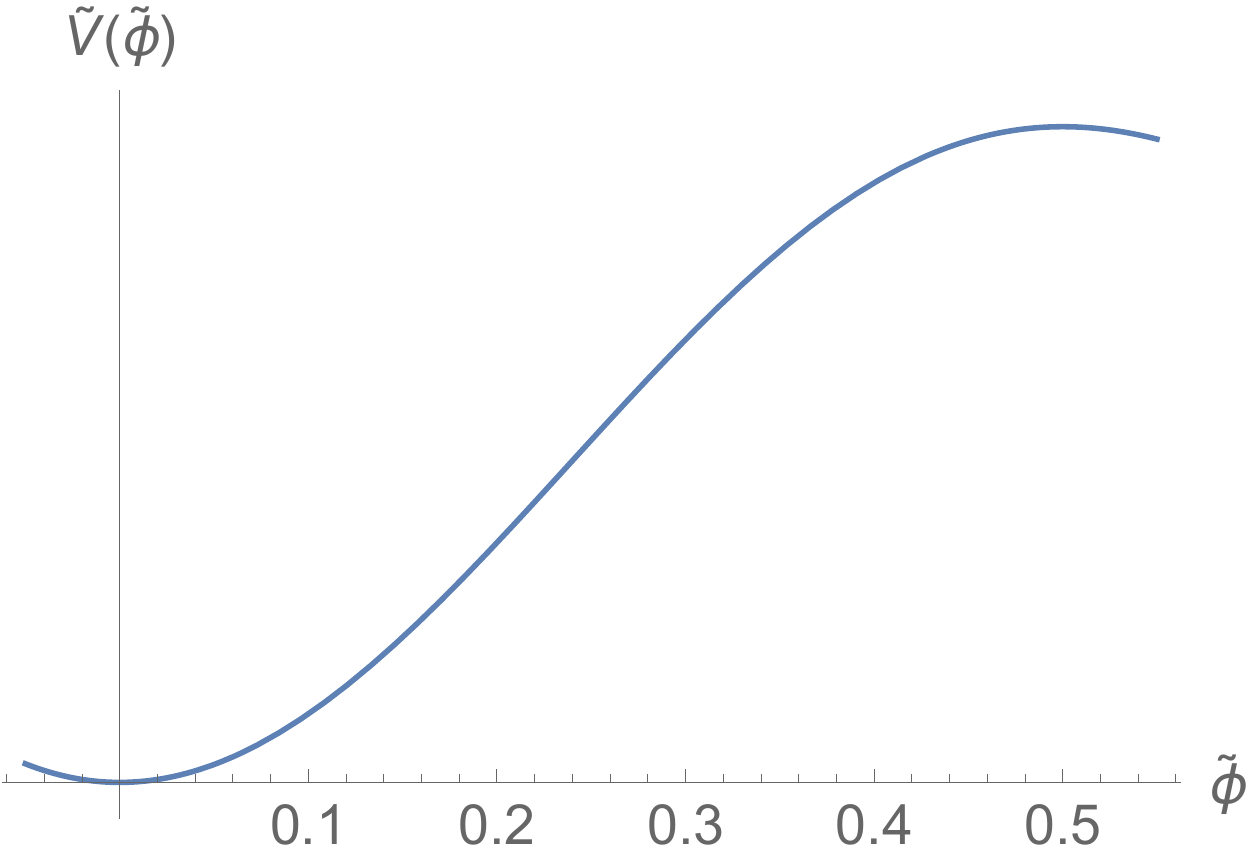}
\caption{~}
\end{subfigure}
\end{center}
\caption{A potential learns to inflate. Beginning with random coefficients (a), the potential flattens into an approximately-linear form (b), before adopting a hilltop shape (c)-(d).}
\label{fig:potentials}
\end{figure}

A typical example of a potential that learns to inflate is shown in Figure \ref{fig:potentials}. The potential is initially drawn at random, which leads to a steep slope and correspondingly a large tensor-to-scalar ratio $r$. As the potential learns to inflate, it flattens out, achieving a flat, approximately-linear shape. Finally, the potential starts to curve, yielding a maximum near the initial value $\tilde \phi_0 = 0.5$. The large, negative second derivative near the maximum yields a negative value for the second slow-roll parameter $\eta$, which in turn leads to a small value of the spectral index $n_s$. Thus, we expect that the typical trajectory of a inflationary model generated by gradient ascent will begin outside the \emph{Planck} 2015 exclusion limits (due to its large value of $r$) and end outside the exclusion limits (due to its small value of $n_s$). The question then becomes, do the trajectories pass through the inclusion regions in the intermediate regime?

\begin{figure}[t!]
\begin{center}
\includegraphics[trim={0cm 0cm 0cm 0cm},clip,scale=0.6]{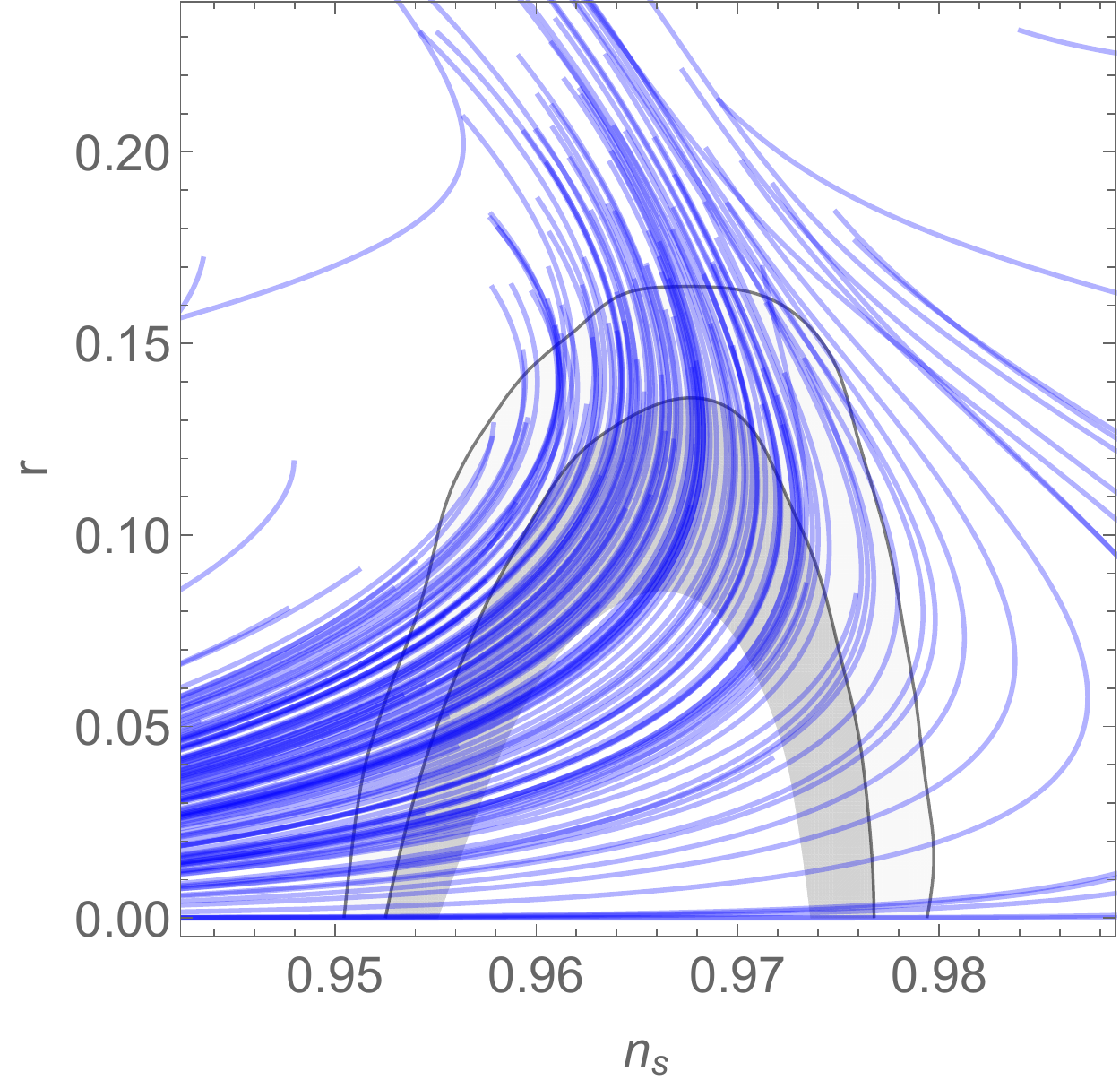}
\end{center}
\caption{Phenomenology of single-field models. Most trajectories begin outside the $1\sigma$, $2\sigma$, and $3\sigma$ exclusion limits of \emph{Planck}, but a sizable fraction enter the inclusion regions as the inflaton learns to inflate and the tensor-to-scalar ratio drops. Eventually, the spectral index becomes too small, and the trajectories exit the inclusion regions.}
\label{fig:trajectories}
\end{figure}

A sample of 200 trajectories in the $n_s$-$r$ plane is shown in figure \ref{fig:trajectories}. It is clear that a large fraction of the trajectories do, in fact, pass through the inclusion regions. In a study of 10,000 trajectories, we found that 64$\%$ of the trajectories passed through the $2\sigma$ inclusion region, while 34$\%$ passed through the $1\sigma$ inclusion region of the \emph{Planck} 2015 data. This agreement with experiment could be interpreted as a success of our gradient ascent approach.

\begin{figure}[t!]
\begin{center}
\includegraphics[trim={0cm 0cm 0cm 0cm},clip,scale=0.6]{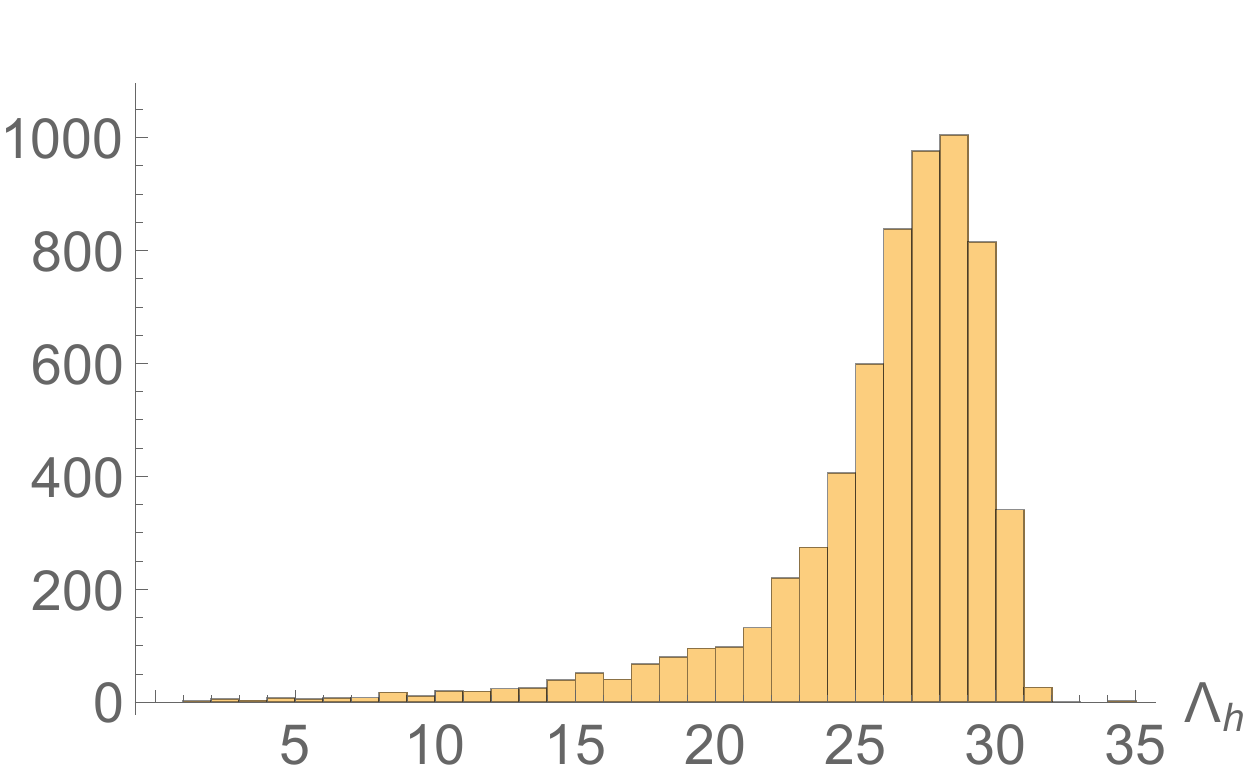}
\end{center}
\caption{Histogram of $\Lambda_h$ values for trajectories exiting the $2\sigma$ inclusion region. Out of 10,000 total trajectories, 6324 enter the $2 \sigma$ exclusion region. Of these, all but 526 exit when $\Lambda_h > 20 M_{\rm{Pl}}$, all but 66 exit when $\Lambda_h > 10 M_{\rm{Pl}}$, and all but two exit when $\Lambda_h > 2 M_{\rm{Pl}}$. Our search produced no models giving viable phenomenology with $\Lambda_h < M_{\rm{Pl}}$.   }
\label{fig:LambdaHist}
\end{figure}

However, viewed from a quantum gravity/effective field theory perspective, these results should be viewed with caution. Figure \ref{fig:LambdaHist} depicts the distribution of the scales $\Lambda_h$ for trajectories to exit the $2 \sigma$ inclusion region. None of our trajectories led to a model with viable phenomenology and a sub-Planckian $\Lambda_h$. Only 66 out of 10,000 trajectories remained within the $2 \sigma$ inclusion region with $\Lambda_h < 10 M_{\rm{Pl}}$, and only two of these 66 remained within the $2 \sigma$ inclusion region with $\Lambda_h < M_{\rm{Pl}}$. The vast majority of the models with $\Lambda_h < 10 M_{\rm Pl}$ have $\eta \sim 1$ and therefore disagree with experiment. This approach leads to a tension between quantum gravity and experiment.

However, if one ignores the complaints of quantum gravity and is willing to blindly consider models with super-Planckian $\Lambda_h$, our gradient ascent approach predicts interesting correlations between cosmic observables. Given the observed limits on the spectral index, one expects a tensor-to-scalar ratio $r \gtrsim 0.05$, which (if correct) will likely be observed within several years by forthcoming CMB experiments \cite{Abazajian:2016yjj}. Assuming a sufficiently-small $r$, our approach favors a redder tilt for the spectral index, $n_s \lesssim 0.965$. Our approach also predicts a field range for phenomenologically-viable inflation, $ \Lambda_h \gtrsim 10 M_{\rm Pl}$.

\begin{figure}[t!]
\begin{center}
\includegraphics[trim={0cm 0cm 0cm 0cm},clip,scale=0.6]{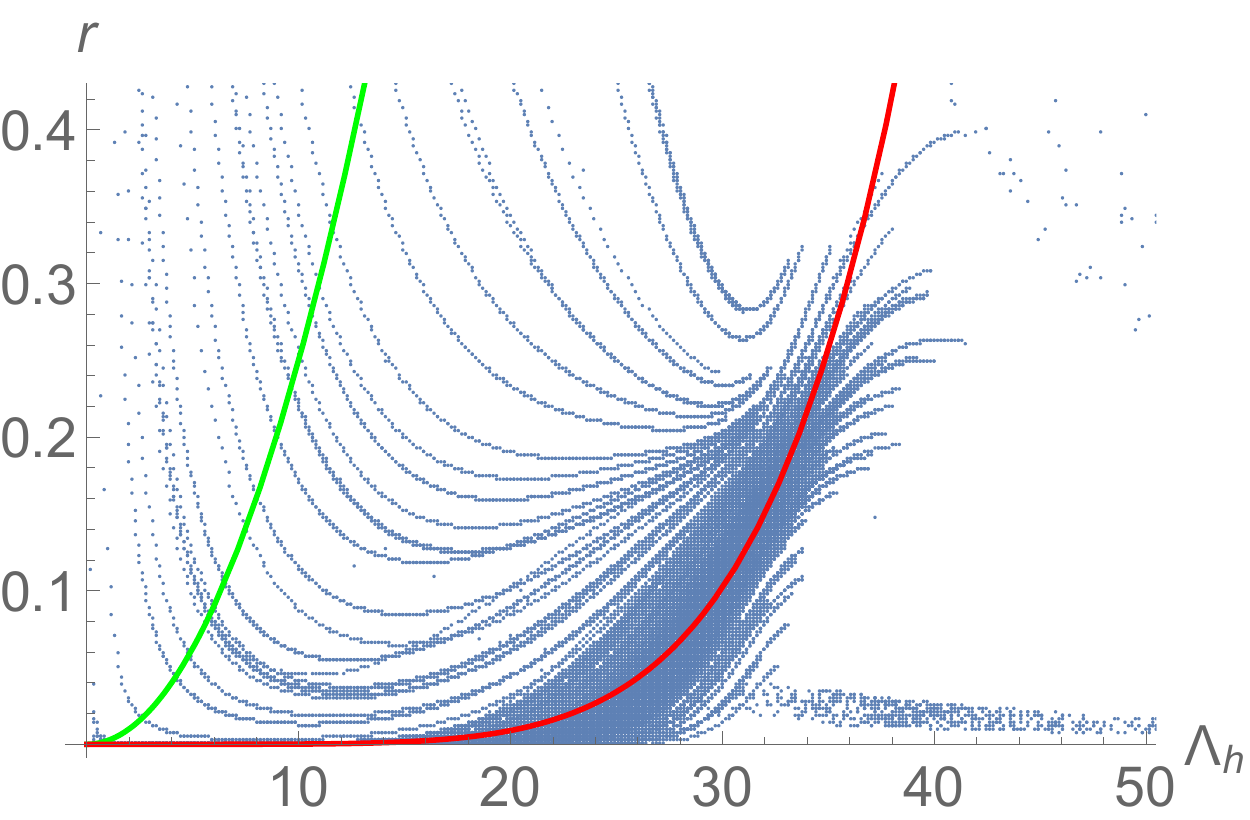}
\end{center}
\caption{$\Lambda_h$ vs. $r$ for a sample of trajectories. The Lyth bound (green) is very conservative, and the typical trajectory instead has $r \sim \Lambda_h^6$ (red). Trajectories that develop an inflection point demonstrate the unusual behavior that $r$ grows as $\Lambda_h$ decreases. }
\label{fig:Lyth}
\end{figure}

\begin{figure}[t!]
\begin{center}
\begin{subfigure}[b]{0.45\textwidth}
\includegraphics[trim={0cm 0cm 0cm 0cm},clip,scale=0.5]{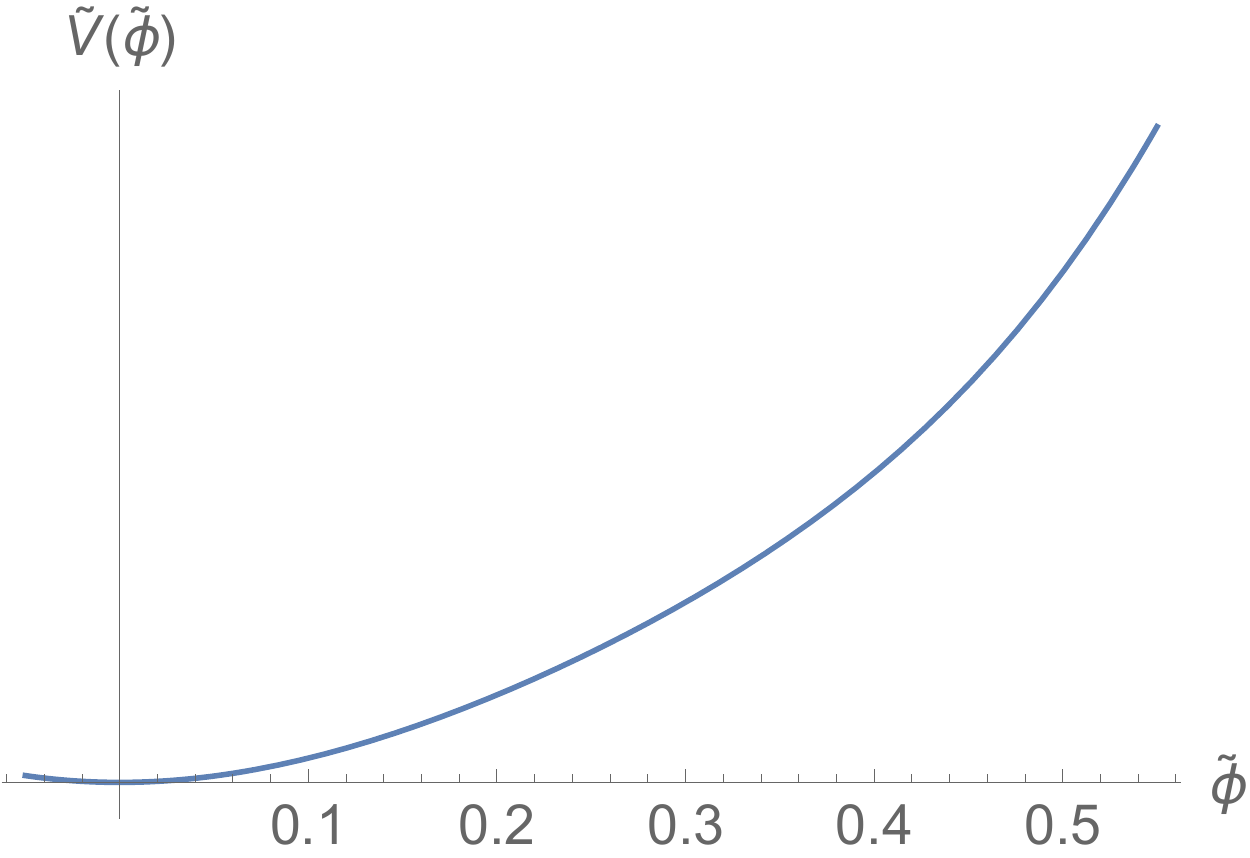}
\caption{~}
\end{subfigure}
~~~~
\begin{subfigure}[b]{0.45\textwidth}
\includegraphics[trim={0cm 0cm 0cm 0cm},clip,scale=0.5]{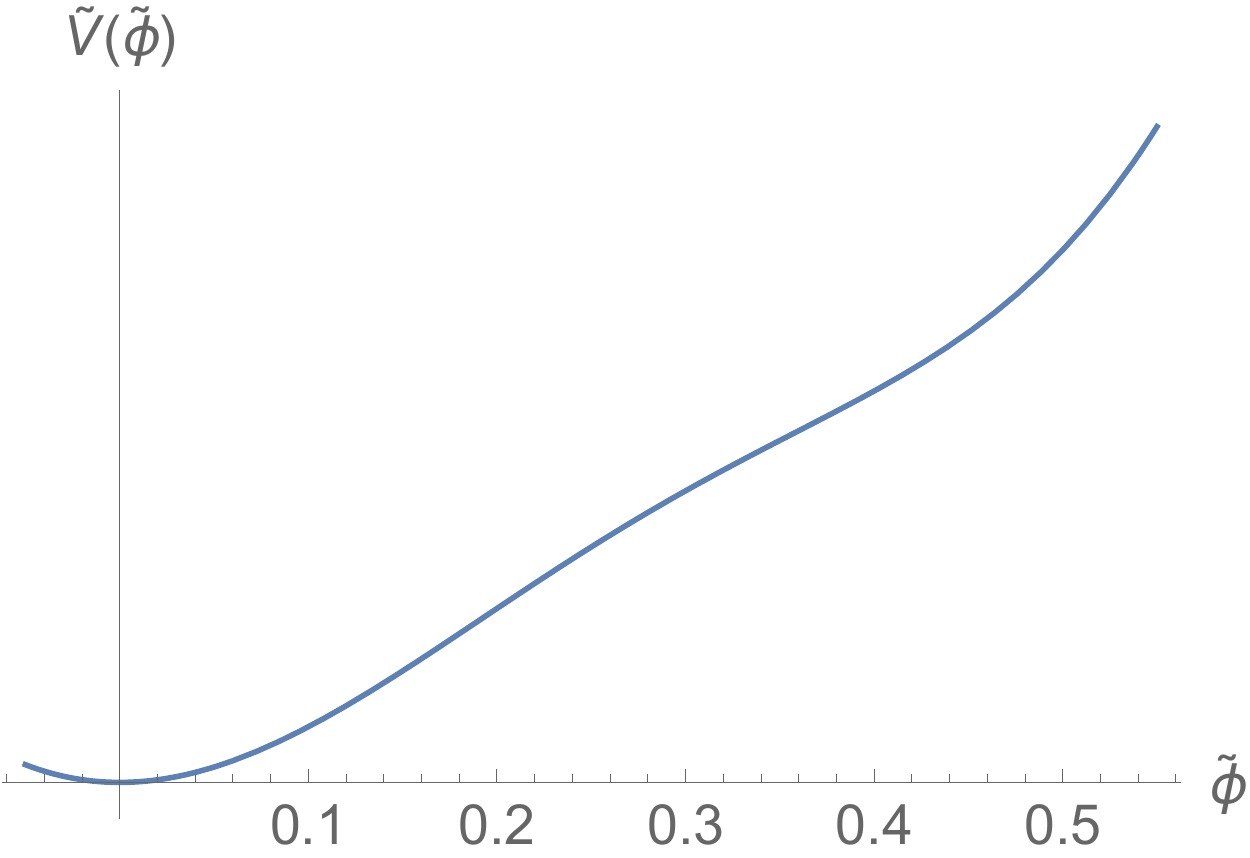}
\caption{~}
\end{subfigure}
\begin{subfigure}[b]{0.45\textwidth}
\includegraphics[trim={0cm 0cm 0cm 0cm},clip,scale=0.5]{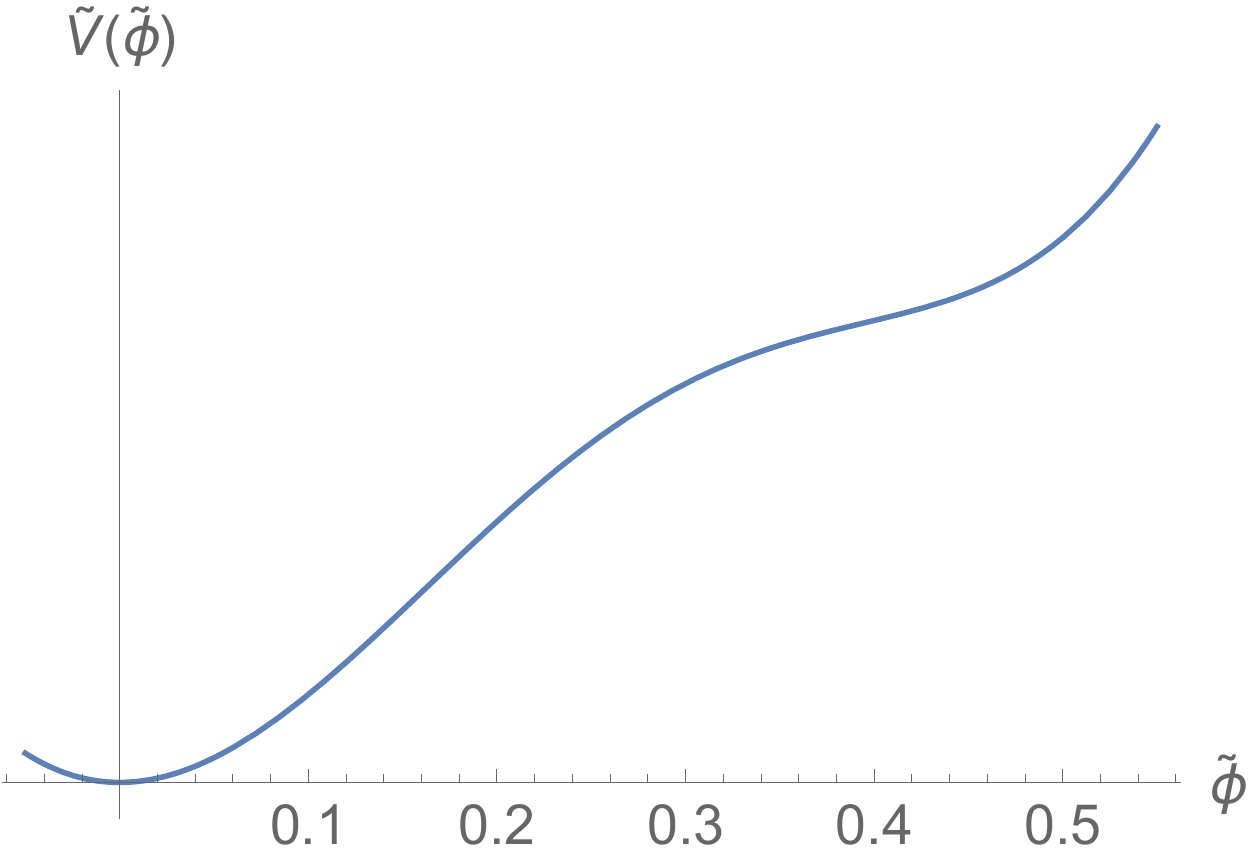}
\caption{~}
\end{subfigure}~~~~
\begin{subfigure}[b]{0.45\textwidth}
\includegraphics[trim={0cm 0cm 0cm 0cm},clip,scale=0.5]{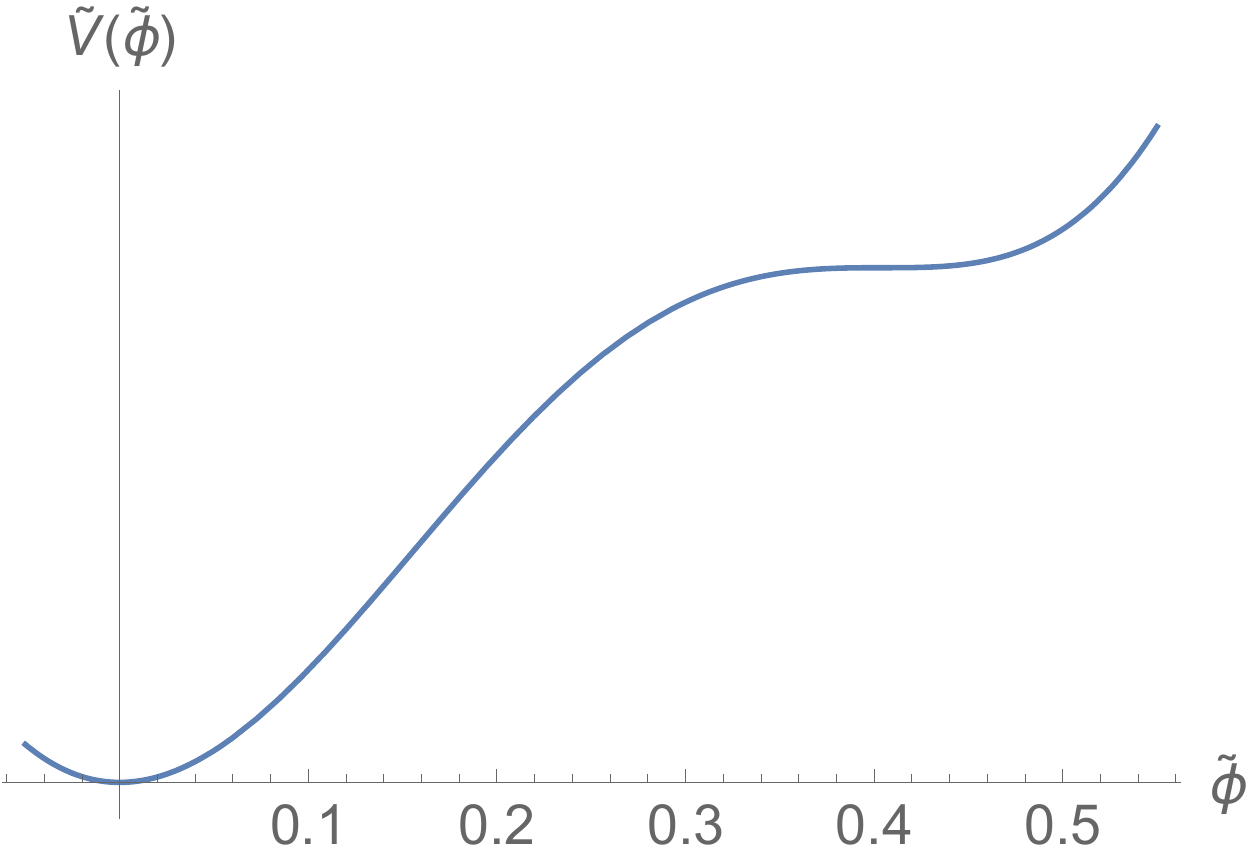}
\caption{~}
\end{subfigure}
\end{center}
\caption{A potential learns to inflate. Beginning with random coefficients (a), the potential eventually develops an inflection point (b)-(d), which yields a significant amount of inflation.}
\label{fig:inflection}
\end{figure}

Using the fact that $\Delta \phi = \Lambda_h \tilde \phi_0 = \Lambda_h /2$, we can also investigate the relationship between $\Delta \phi$ and the tensor-to-scalar ratio in our gradient ascent approach. While the Lyth estimate \cite{Lyth:1996im} gives
\begin{equation}
\Lambda_h \gtrsim 2 \left( \frac{r}{0.01} \right)^{1/2},
\end{equation}
we find that this bound is very conservative, and indeed our hilltop trajectories are better approximated by 
\begin{equation}
\Lambda_h \sim 20.4 \times \left( \frac{r}{0.01} \right)^{1/6},
\end{equation}
as shown in figure \ref{fig:Lyth}. One notes that some trajectories in this figure do not fit this curve at all and instead tend to give larger $r$ at smaller $\Lambda_h$! These bizarre trajectories correspond to the inflection point models of figure \ref{fig:inflection} rather than the hilltop models of figure \ref{fig:potentials}. These inflection point models tend to give $n_s \gg 1$ under this approach, so they are uninteresting from the perspective of phenomenology.

However, the strange behavior of inflection point models seen here is an artifact of this first approach, in which we fix $N_e = 60$ and use this to determine $\Lambda_h$. In these models, this means that almost all of the inflation is occurring along the plateau near the inflection point, but the phenomenology is being computed using the value of the slow-roll parameters at $\tilde \phi_0=0.5$, where the first derivative of the potential is large (see figure \ref{fig:inflection} (c)-(d)). This represents a rather drastic fine-tuning of these models: $\Lambda_h$ has been tuned just right so that the part of the potential relevant for phenomenology represents a very tiny fraction of the total number of $e$-foldings produced by the potential. If $\Lambda_h$ were decreased slightly, this potential would no longer produce enough inflation. If $\Lambda_h$ were increased slightly, the modes relevant for phenomenology would exit the horizon when the inflaton was rolling along the plateau, and the phenomenology would be very different. We see that while our approach of fixing $N_e = 60$ and varying $\Lambda_h$ yields sensible results in the $90\%$ of trials that produced a local maximum near $\tilde \phi_0 = 0.5$, it leads to issues in the $10\%$ of trials that produced inflection points. We therefore are lead to consider a second approach: fixing $\Lambda_h$, and letting $N_e$ vary. We will see that this not only remedies the bizarre dependence of $r$ on $\Lambda_h$, but also generates viable models of small-field inflation with $\Lambda_h < M_{\rm Pl}$.

\subsection{Small-Field Inflation at Fixed $\Lambda_h$, Variable $N_e$}\label{ssec:small}

We now restrict our analysis to the ``problematic" trajectories--namely, the ones that produce an inflection point rather than a local maximum. We saw that our previous approach led to problems due to an implicit fine-tuning of $\Lambda_h$. To remedy this problem, we fix $\Lambda_h$ and allow the total number of $e$-folds $N_e$ to vary, though we still compute phenomenology at a pivot scale 60 $e$-folds before the end of inflation. As a consequence, the value $\tilde \phi_*$ at which phenomenology is determined is no longer fixed to be $\tilde \phi = 0.5$, but instead can vary between $0$ and $0.5$ (assuming that $\Lambda_h$ is chosen sufficiently large so that the model produces at least 60 $e$-folds in total). Since most of the inflation occurs close to the inflection point in our models, $\tilde \phi_*$ will typically be very close to the inflection point. This alleviates the fine-tuning discussed above, which put $\tilde \phi_* = \tilde \phi_0 = 0.5$ regardless of the position of the inflection point.

How does gradient ascent act on a model with an inflection point? To good approximation, we can approximate the potential in the vicinity of an inflection point $\tilde \phi_{\rm inf}$ as a cubic,
\begin{equation}
\tilde V(\tilde \phi) = W_0 + W_1 (\tilde \phi - \tpi) + \frac{1}{3!} W_3 (\tp - \tpi)^3.
\end{equation}
Successful inflation requires $W_1 W_3 /W_0^2 \ll \Lambda_h^4$. Given such a model, gradient ascent effectively decreases $W_1$ indefinitely without significantly altering $W_0$ or $W_3$. Thus, given a model with set $W_0$, $W_3$, we can approximate the effect of gradient ascent to good accuracy by merely letting $W_1 \rightarrow 0$. For any $\Lambda_h$ fixed, no matter how small, this produces a model with arbitrarily-large number of $e$-folds.
 
 The phenomenology of cubic inflection models has been studied extensively \cite{Baumann:2007ah, Linde:2007jn,Masoumi:2016eag}, and we can make use of these results here. In particular, we may express the total number of $e$-folds produced near an inflection point as
 \begin{equation}
 N_{\rm{tot}} := \Lambda_h^2 \tilde N_{\rm{tot}} \approx \pi \sqrt{2} \frac{W_0}{\sqrt{W_1 W_3}} \Lambda_h^2.
 \label{eq:infNtot}
 \end{equation}
Then, we have
\begin{equation}
n_s \approx 1 - \frac{4 \pi}{ N_{\rm{tot}}} \cot \left( \frac{\pi N_e}{ N_{\rm{tot}}} \right),
\label{eq:infns}
\end{equation}
where we set $N_e = 60$. For small-field inflation, $\Lambda_h \lesssim 1$, the tensor-to-scalar ratio is too small to be detectable, so we may simply focus on the spectral index. From here, we see that our inflection point models exhibit a universal behavior in that the spectral index depends only on the total number of $e$-folds produced by the model, and in turn the total number of $e$-folds depends only on the ratio $W_0^2 \Lambda_h^4/(W_1 W_3)$ appearing in (\ref{eq:infNtot}). Since gradient acts to simply decrease $W_1$, thereby increasing $N_{\rm{tot}}$, we see that $n_s$ will begin above the allowed range, pass through the inclusion region, and then decrease below it. In the single-field case, we see that our approach therefore makes no meaningful predictions for $n_s$ beyond the predictions of a generic inflection point model, since any physical information contained in $W_0$, $W_3$, and $\Lambda_h$ can be absorbed into $W_1$, which depends only on the number of steps performed in the gradient ascent. Nonetheless, gradient ascent can still be a useful tool for producing such inflection point models, so it is useful to study the behavior of our model parameters from a practical perspective.

Using (\ref{eq:infns}), we see that a phenomenologically-viable $n_s$ requires $144 \lesssim N_{\rm{tot}}\lesssim 211$. Among the gradient ascent trajectories that produced an inflection point, we find that the ratio $W_0^2/W_3$ is given by
\begin{equation}
\frac{W_0^2}{W_3} \approx 10^{-2.4 \pm 0.4},
\end{equation}
with $W_0$ typically $O(10^0)$ and $W_3$ typically $O(10^3)$. Thus, for a model that produces a phenomenologically-viable $n_s$, the value of $W_1$ is bounded below as
\begin{equation}
W_1 \gtrsim 10^{-5.4 \pm 0.4} \Lambda_h^4.
\end{equation}
We see that for $\Lambda_h \lesssim 1$, $W_1$ must be $O(10^{-5})$, with $W_1$ decreasing drastically as $\Lambda_h$ decreases. This leads to practical issues with the gradient ascent. Namely, if the learning rate is too small, it can take a very long time for the inflection point to develop and $W_1$ to decrease to the point required for acceptable phenomenology. If the learning rate is too large, the model will overshoot the ``sweet spot" with $144 \lesssim N_{\rm{tot}}\lesssim 211$, and if it overshoots far enough, will even produce a model with $W_1 < 0$, which is not monotonic on the full range $0 < \tilde \phi < \tilde \phi_0$. Indeed, for the learning rate used in our analysis, setting $\Lambda_h = 1$, around $20\%$ of the trajectories overshot the sweet spot entirely, and only $10 \%$ of the trajectories included more than one point in the sweet spot. A better choice for the learning rate for an inflection point model would ensure that the step size is proportional to $W_1$, preventing $W_1$ from overshooting the maximum at $W_1 = 0$.

A similar statement applies to our hilltop models. In that case, the maximum of $\tilde N_e$ occurs not as $W_1 \rightarrow 0$, but rather as the position $\tilde \phi_{\rm{max}}$ of the maximum tends towards $\tilde \phi_0 = 0.5$. Overshooting the maximum $\tilde \phi_{\rm{max}} = \tilde \phi_0$, which produces an infinite number of $e$-folds, is very common. A better choice of learning rate would ensure that the step size is proportional to $\tilde \phi_{\rm{max}} - \tilde \phi_0$. However, such small-field hilltop models are less interesting than inflection point models from the perspective of phenomenology, as they yield a value of $n_s$ that is too small to agree with experiment \cite{Linde:2007jn}.

We have seen that our approaches lead to very different results depending on whether the potential develops a local maximum or an inflection point. It is therefore important to determine whether gradient ascent from some initial $c_i^{(0)}$ will produce a local maximum or an inflection point. Supervised machine learning is well-suited to this sort of task. However, in our case, we are not merely interested in making predictions but also in developing intuition. A simple logistic regression works well for this purpose:
\begin{equation}
p(\{ c_i^{(0)} \} ) = \left[ 1+ \exp (- \beta - \sum_{i=2}^{N}  w_i c_i^{(0)}) \right]^{-1}
\end{equation}
and $y(\{ c_i^{(0)} \}) = 0$ if gradient ascent produces a local maximum and $y(\{ c_i^{(0)} \}) = 1$ if gradient ascent produces an inflection point. We find that this method of simple logistic regression is capable of predicting whether an initial vector of coefficients $\{ c_i^{(0)} \}$ will yield an inflection point or a local maximum with an accuracy of about $85\%$. The model takes the form
\begin{align}
\beta &= 1168, w_2=-40.3,  w_3=-29.9,  w_4=-3.0,  w_5=12.0,  w_6=17.8,  w_7=13.2,  w_8=10.8,  \nonumber \\ w_9&=7.6,  w_{10}=6.8,  w_{11}=3.7,  w_{12}=1.5, w_{13}=1.4, w_{14}=1.5, w_{15}=-0.9.
\end{align}
Of these, only $\beta$ and $w_i, 2 \leq i \leq 11$ are statistically significant. When the linear combination $ \beta + \sum_{i} w_i c_i^{(0)}$ is positive (negative), the model predicts an inflection point (local maximum). We see that the key to developing an inflection point is large, positive, initial values of $c_i, i \geq 5$. These lead to higher-order terms in the potential that continue to grow as the model learns to inflate, leading to an inflection point. On the other hand, if the higher-order terms in the potential are negative, or if the potential is dominated by lower-order terms (i.e. if $c_i, i \leq 4$ are initially large), then the potential typically develops a local maximum. It is interesting that the coefficients with positive weights $w_i$--namely $c_i^{(0)}, i \geq 5$--are precisely those that are suppressed by powers of the UV scale $\Lambda$ in effective field theory. But (especially given our distinction between the scales $\Lambda_v$ and $\Lambda_h$) it is unclear to us if there is any deeper reason why this should be the case.

\subsection{Two-Field Inflation}

\begin{figure}[t!]
\begin{center}
\begin{subfigure}[b]{0.45\textwidth}
\includegraphics[trim={0cm 0cm 0cm 0cm},clip,scale=0.5]{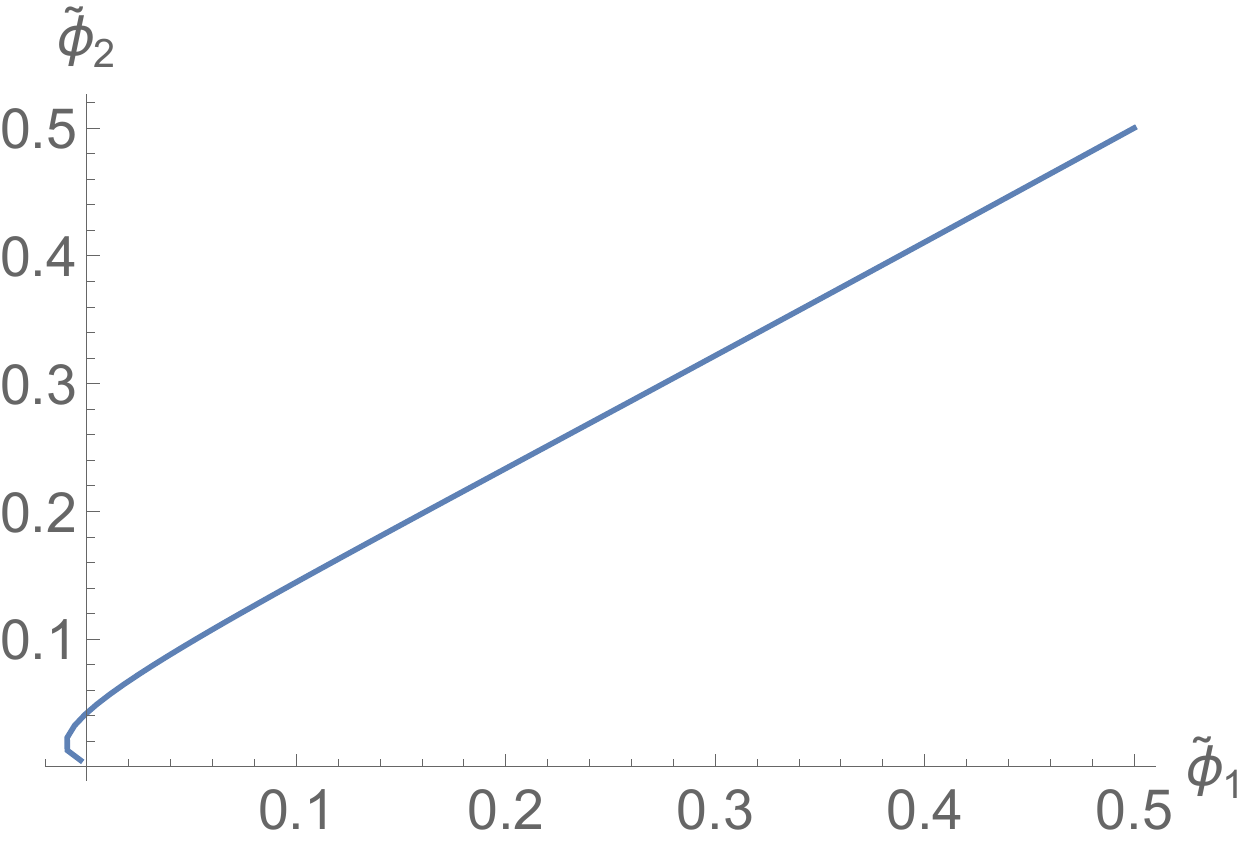}
\caption{~}
\end{subfigure}
~~~~
\begin{subfigure}[b]{0.45\textwidth}
\includegraphics[trim={0cm 0cm 0cm 0cm},clip,scale=0.5]{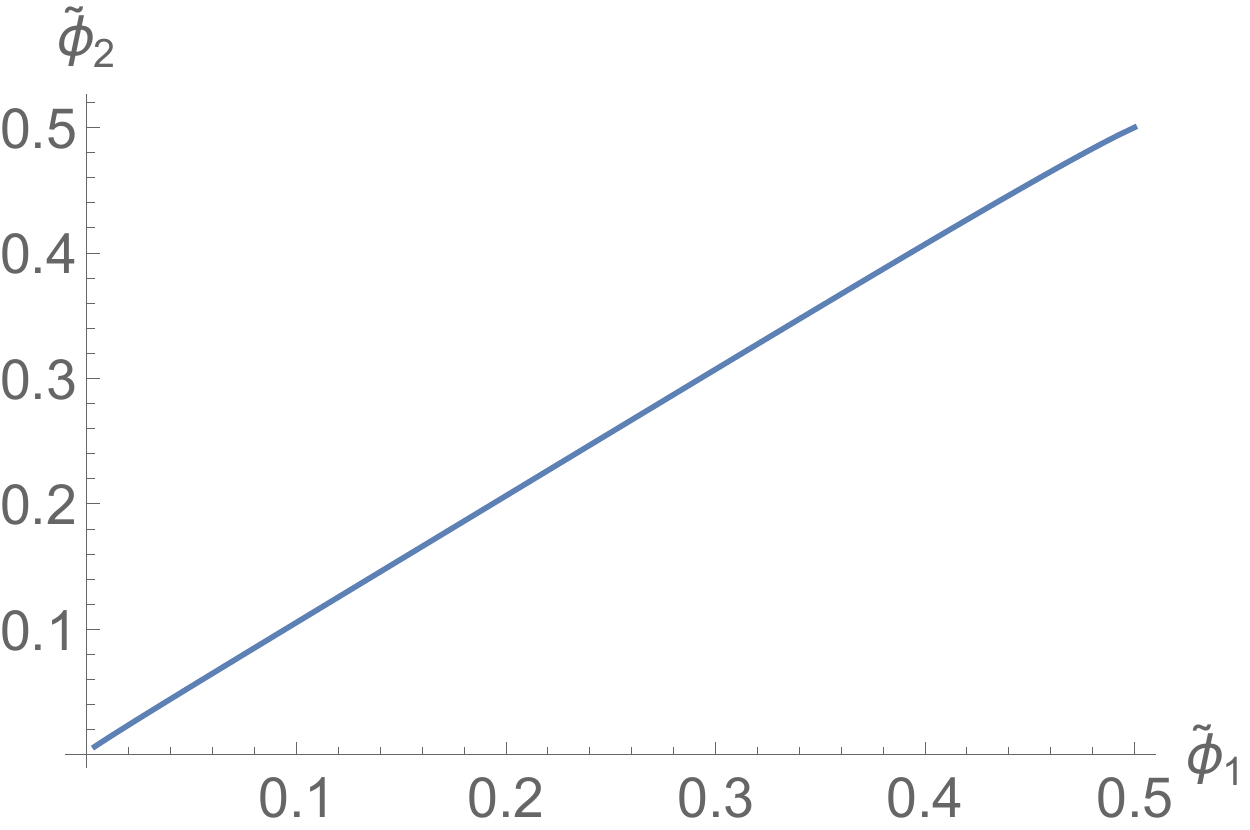}
\caption{~}
\end{subfigure}
\begin{subfigure}[b]{0.45\textwidth}
\includegraphics[trim={0cm 0cm 0cm 0cm},clip,scale=0.5]{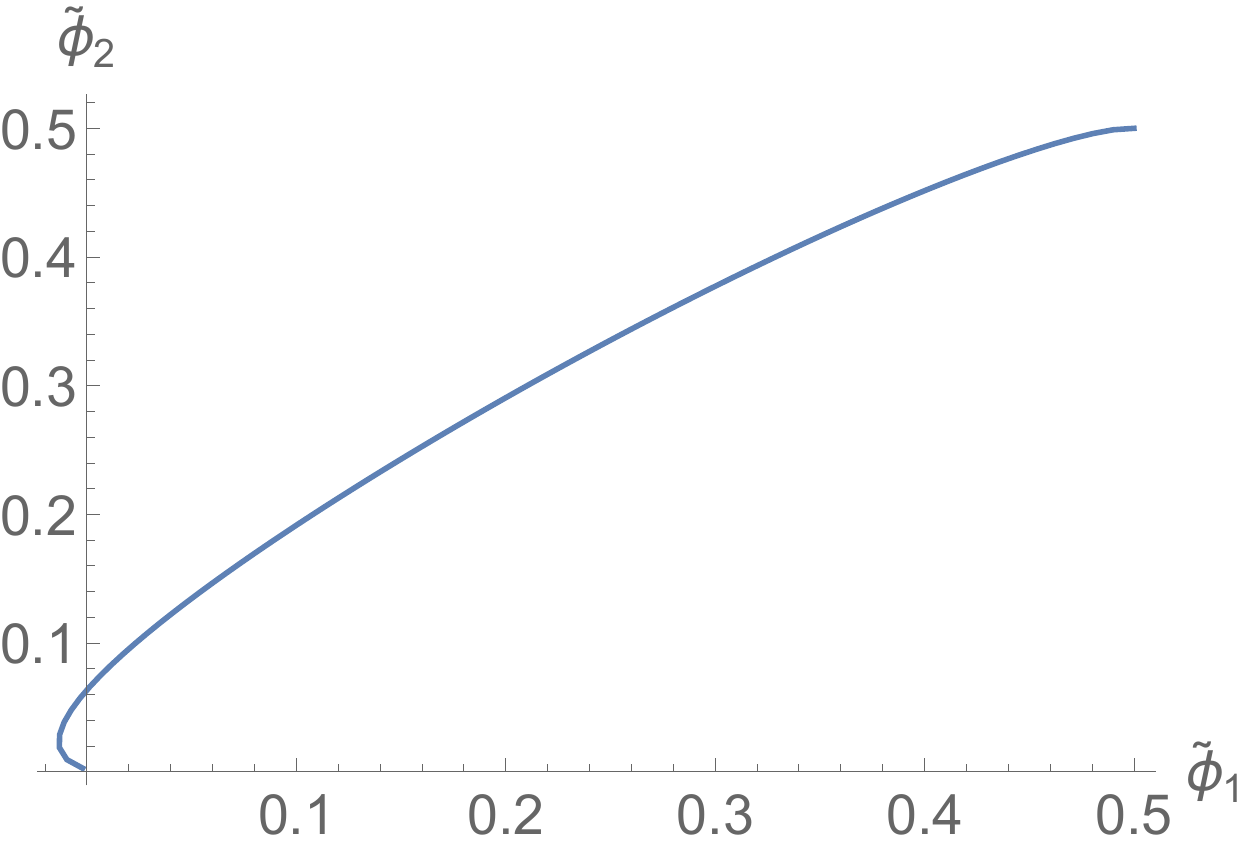}
\caption{~}
\end{subfigure}~~~~
\begin{subfigure}[b]{0.45\textwidth}
\includegraphics[trim={0cm 0cm 0cm 0cm},clip,scale=0.5]{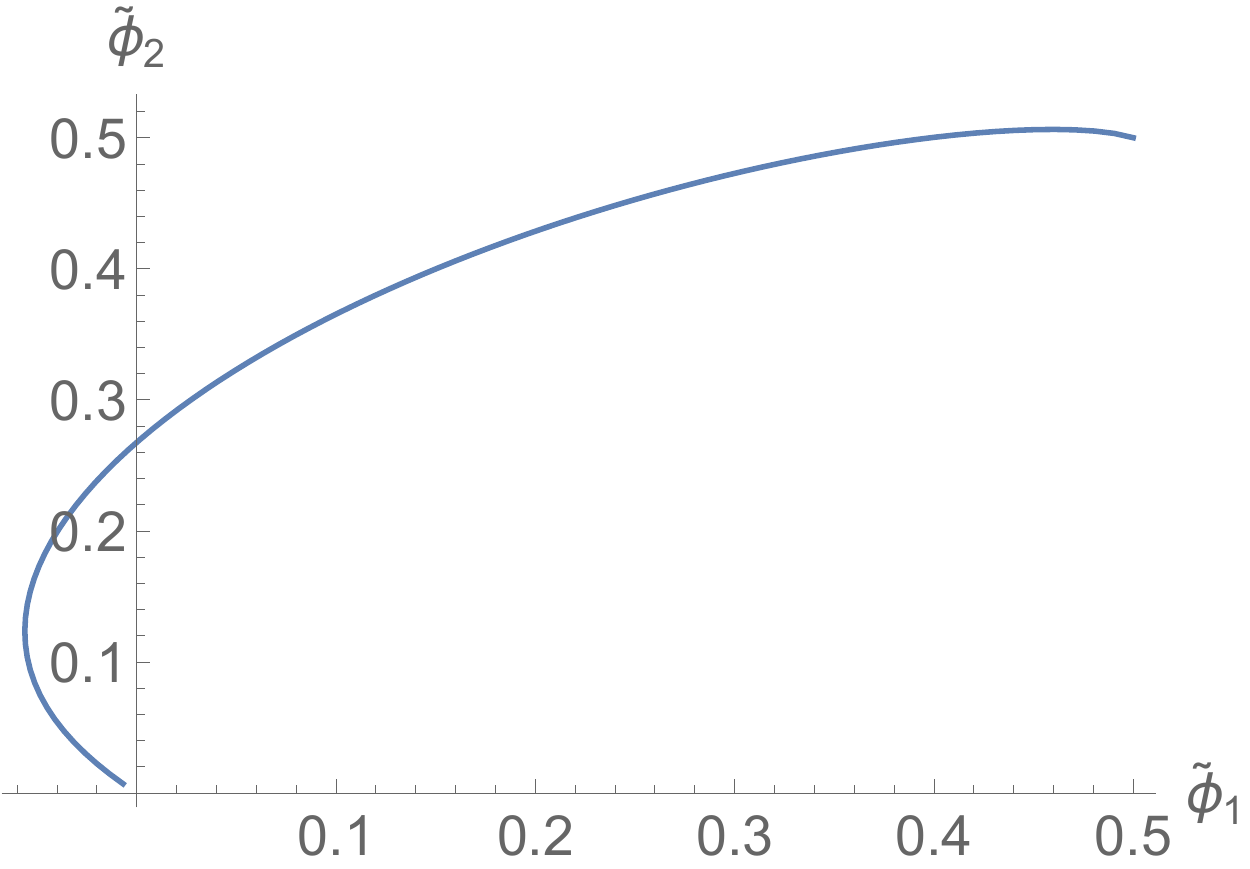}
\caption{~}
\end{subfigure}
\end{center}
\caption{The inflaton path of a two-field model that learns to inflate. Initially, the path features a sharp bend, indicating a hierarchy in the masses of the two fields (a). Next, the inflaton path becomes effectively single-field, as the hierarchy disappears (b). Eventually, however, the path becomes curved again, and the potential yields a genuine multi-field model of inflation (c)-(d).}
\label{fig:paths}
\end{figure}

The typical evolution of the inflaton path for a two-field model that learns to inflate is shown in figure \ref{fig:paths}. An initial hierarchy in the masses of the two fields leads to a sharp bend in the inflaton path. However, the model quickly learns ``isotropic $N$-flation" \cite{liddle:1998jc, dimopoulos:2005ac}--the hierarchy disappears, and the inflaton learns to inflate along the diagonal of two nearly-identical fields, yielding an effectively single-field model.

\begin{figure}[t!]
\begin{center}
\begin{subfigure}[b]{0.45\textwidth}
\includegraphics[trim={0cm 0cm 0cm 0cm},clip,scale=0.5]{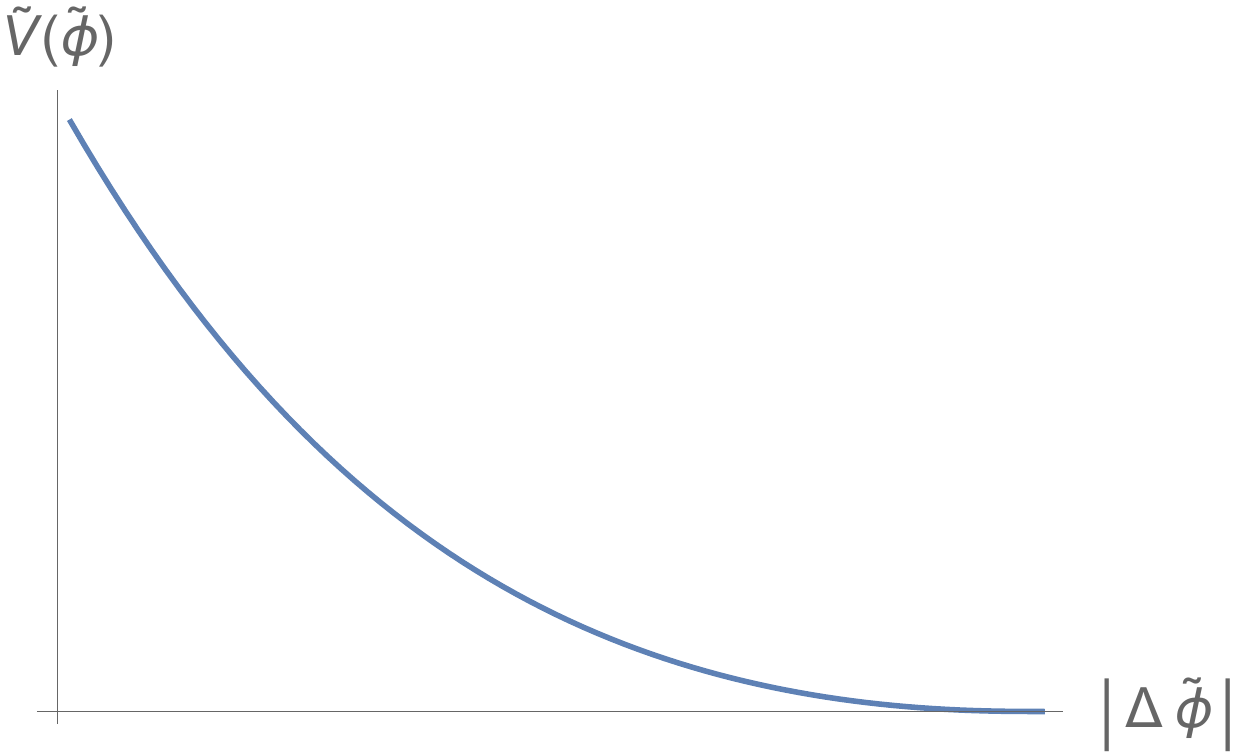}
\caption{~}
\end{subfigure}
~~~~
\begin{subfigure}[b]{0.45\textwidth}
\includegraphics[trim={0cm 0cm 0cm 0cm},clip,scale=0.5]{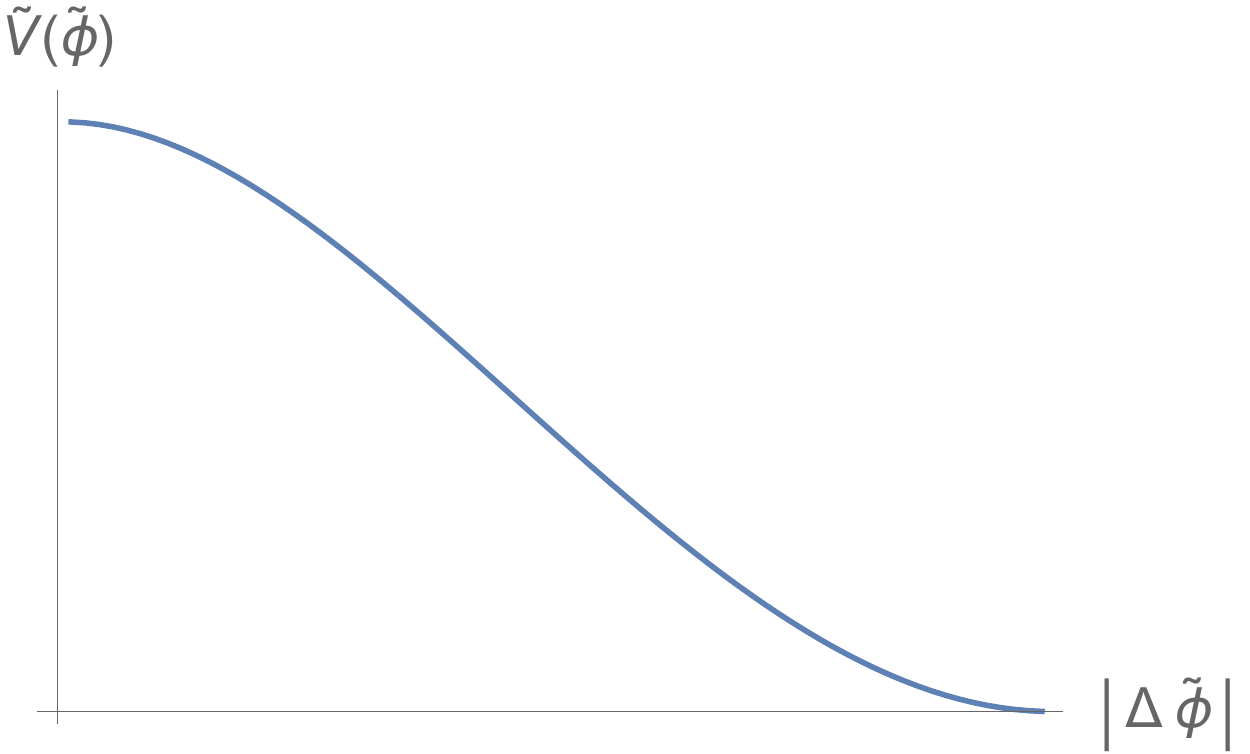}
\caption{~}
\end{subfigure}
\begin{subfigure}[b]{0.45\textwidth}
\includegraphics[trim={0cm 0cm 0cm 0cm},clip,scale=0.5]{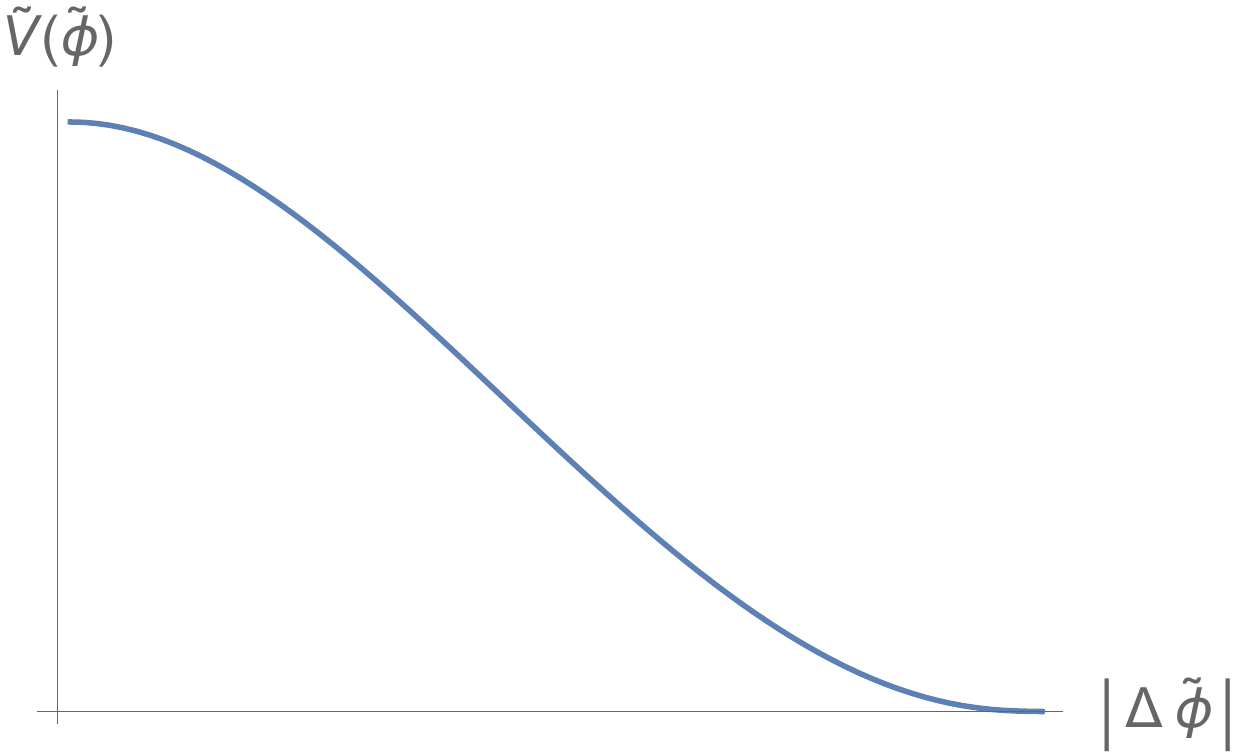}
\caption{~}
\end{subfigure}~~~~
\begin{subfigure}[b]{0.45\textwidth}
\includegraphics[trim={0cm 0cm 0cm 0cm},clip,scale=0.5]{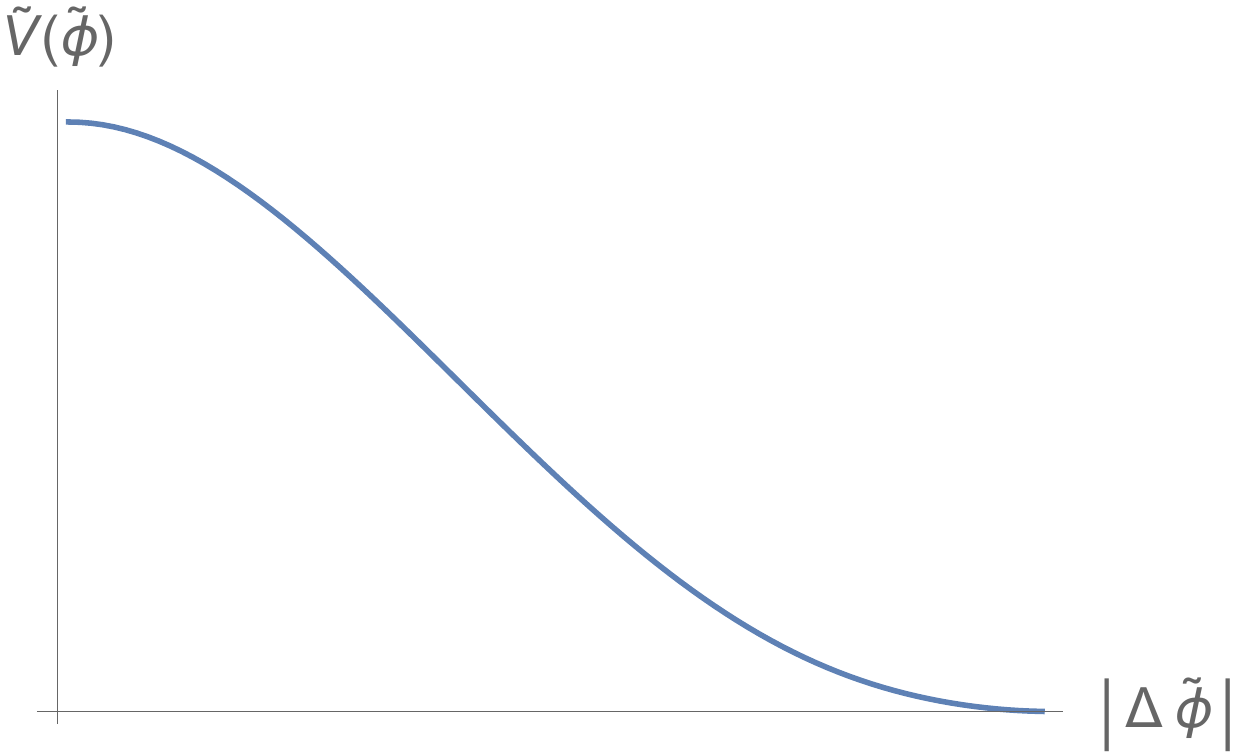}
\caption{~}
\end{subfigure}
\end{center}
\caption{The potential along the paths of the two-field models in figure \ref{fig:paths}. The potential is initially very steep (a) but adopts a hilltop while the model is still effectively single-field (b). This hilltop persists as the path becomes curved (c)-(d).}
\label{fig:Vtwofield}
\end{figure}

One might have expected the model to remain as a single-field model, eventually adopting a hilltop shape as we saw in the single-field case. The model does in fact develop a hilltop shape (see figure \ref{fig:Vtwofield}), but perhaps surprisingly, the dynamics of the model eventually become multi-field! As shown in figure \ref{fig:paths} (d), the inflaton path becomes significantly curved, and $\eta_\perp/v $ grows larger than 1, violating the slow-turn approximation (see appendix \ref{sec:twofieldapp}). Presumably, this is due to the fact that a curved path is longer than a straight path, so more $e$-folds of inflation can be generated by the former than the latter. Note that if the potential were exactly symmetric in $\phi^1$ and $\phi^2$, there would be no reason for gradient ascent to break the symmetry, so the model would remain symmetric forever. Thus, models of isotropic $N$-flation lie along the stable direction of a hilltop-like saddle point of the function $\tilde N_e(\{ c_{jm} \})$.

Shortly after the inflaton path develops significant curvature, it tends to stray too far from the origin, and the inflaton falls into a different local minimum. At this point, our gradient ascent terminates.

\begin{figure}[t!]
\begin{center}
\includegraphics[trim={0cm 0cm 0cm 0cm},clip,scale=0.6]{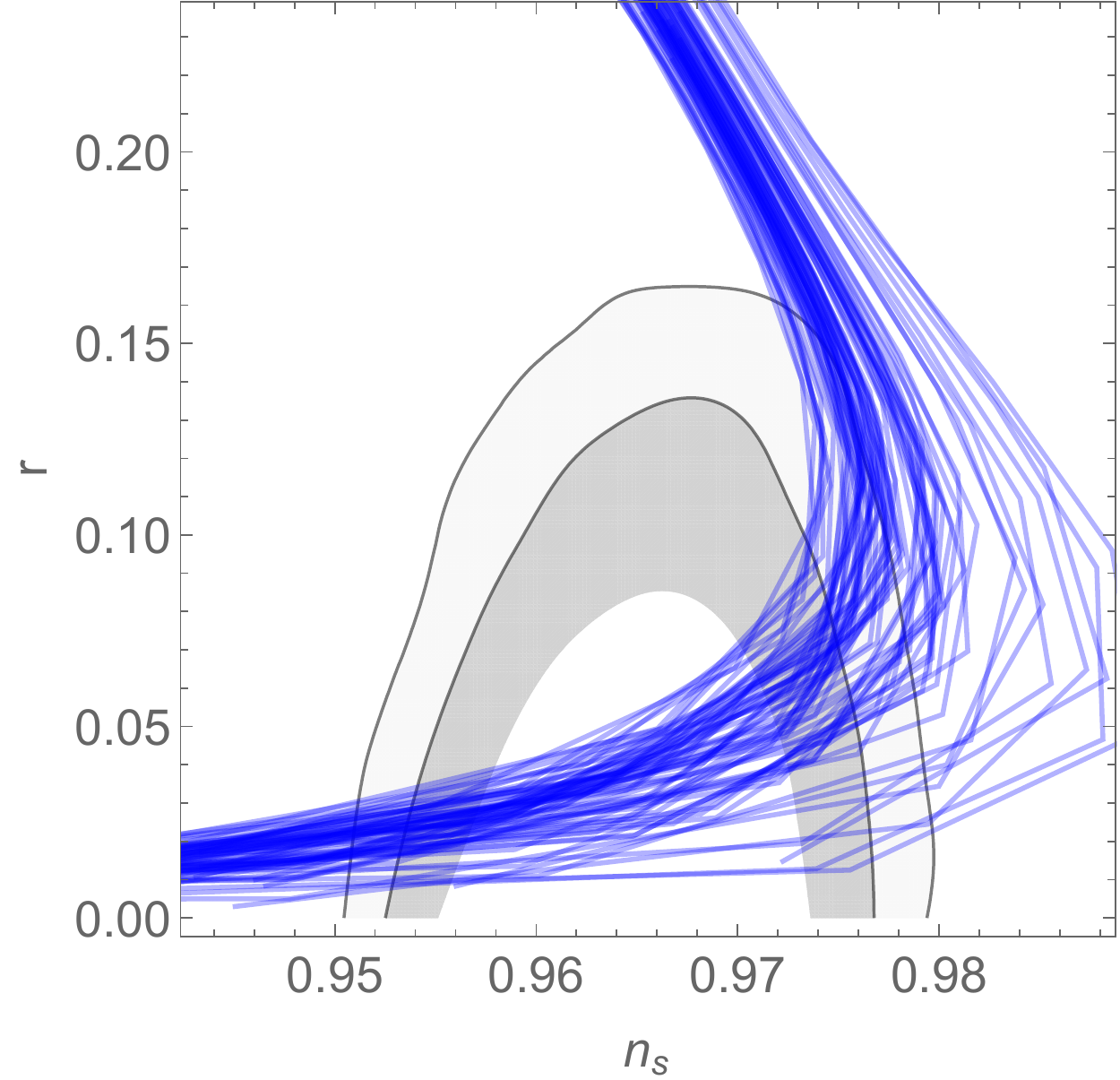}
\end{center}
\caption{Phenomenology of two-field models. The $n_s$ values for a given $r$ are slightly larger than those of the single-field models.}
\label{fig:twofieldnsr}
\end{figure}

The phenomenology of most of the two-field models along the gradient ascent trajectory does not differ qualitatively from that of the single-field models already considered, as shown in Figure \ref{fig:twofieldnsr}. Initially, the models tend to have large $r$, but as the model learns to inflate, $r$ and $n_s$ both decrease. In some intermediate regime, the model falls inside the $2\sigma$ \emph{Planck} inclusion region. These models do not lead to significant non-Gaussianity, and they require super-Planckian $\Lambda_h$.

There are several differences between the single-field and two-field results, however. First, our two-field models tend to have a slightly larger $n_s$, which brings them into better agreement with experiment. Indeed, over 97\% of our trajectories passed through the $2\sigma$ \emph{Planck} inclusion region, and 72\% of trajectories passed through the $1\sigma$ inclusion region.

\begin{figure}[t!]
\begin{center}
\begin{subfigure}[b]{0.45\textwidth}
\includegraphics[trim={0cm 0cm 0cm 0cm},clip,scale=0.5]{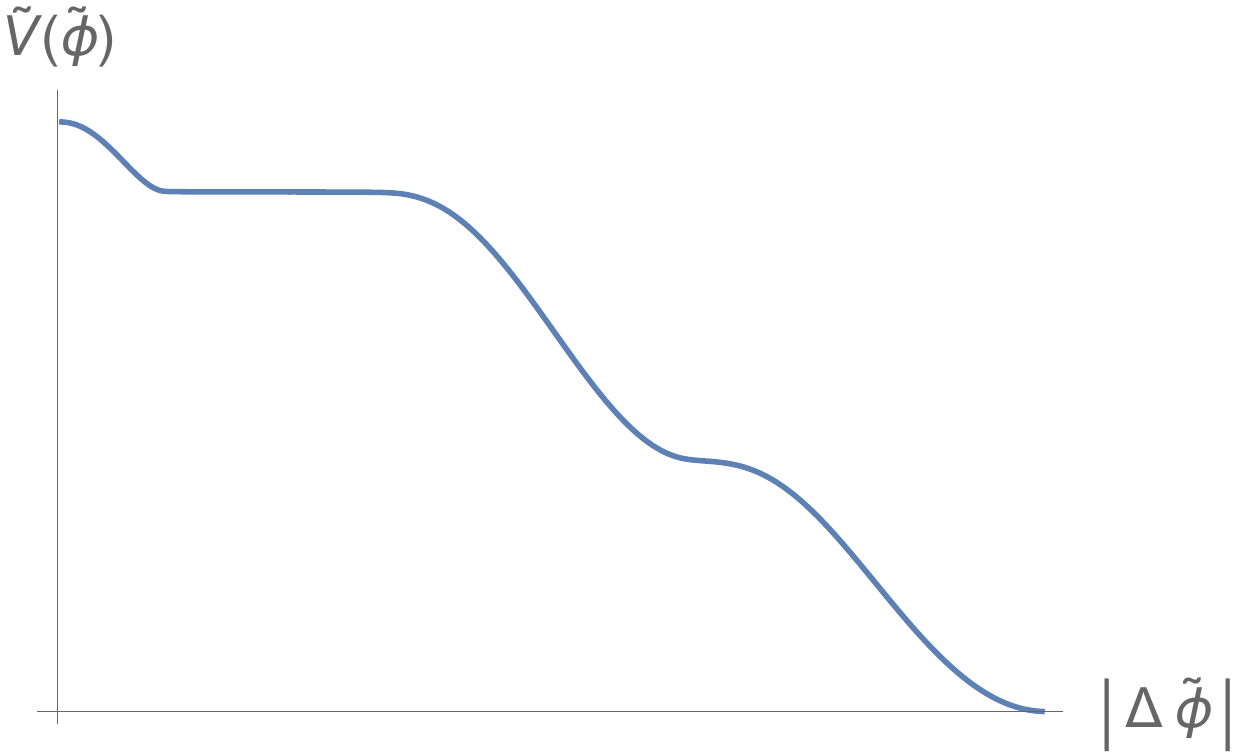}
\end{subfigure}
~~~~
\begin{subfigure}[b]{0.45\textwidth}
\includegraphics[trim={0cm 0cm 0cm 0cm},clip,scale=0.5]{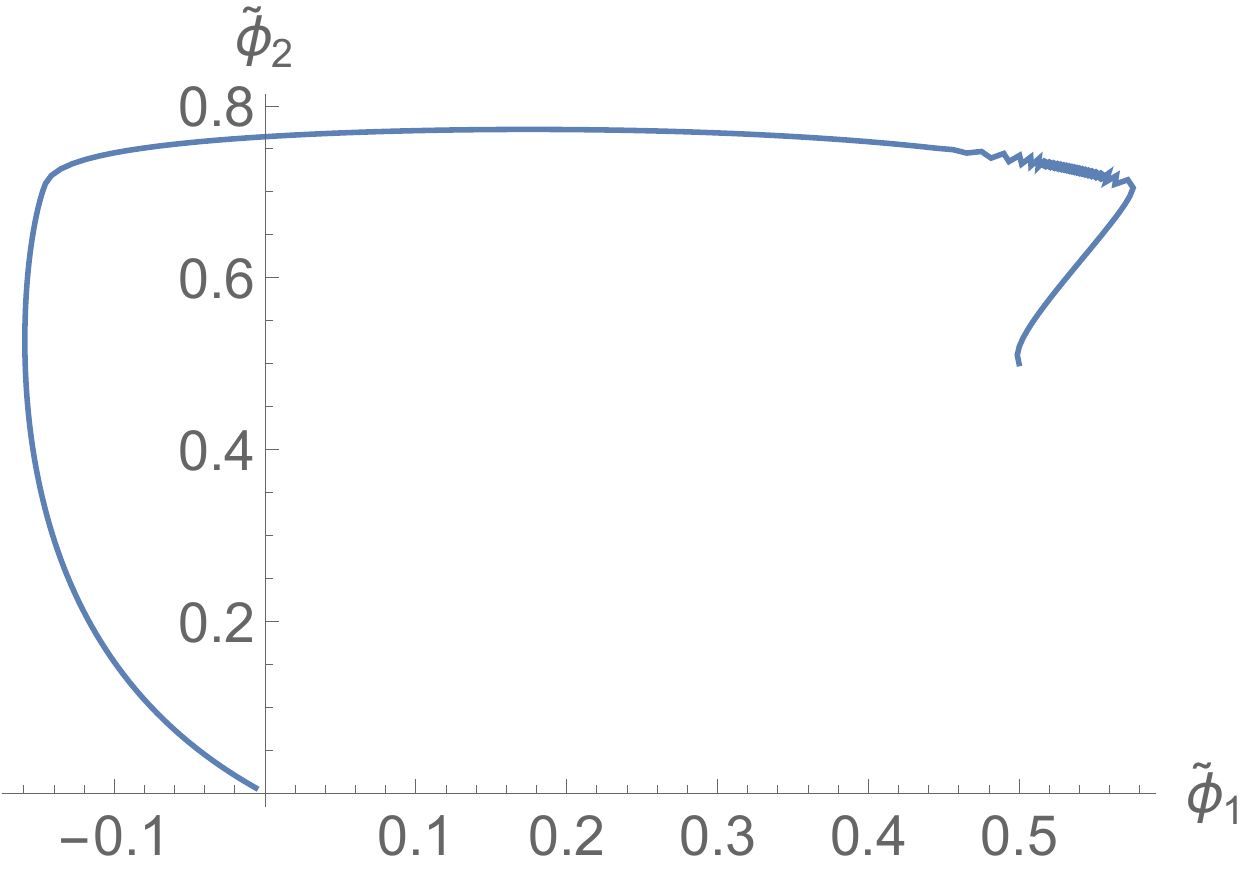}
\end{subfigure}
\end{center}
\caption{A two-field inflaton path with multiple plateaus. The plot on the left shows the potential of the path as a function of distance traveled, while the one on the right depicts the path in field space. }
\label{fig:bizarre}
\end{figure}

Second, our two-field models can show more complicated limiting behavior than simple local maxima or inflection points. Figure \ref{fig:bizarre} shows an example of a two-field path with \emph{three} plateaus (including the short one near ${ \boldsymbol{ \tilde \phi}}_0$). This sort of behavior is quite rare, however, appearing in less that $5\%$ of our trials. Simple saddle points/local maxima are much more common.

Finally, as previously discussed, our two-field models eventually exhibit multi-field dynamics, as the inflaton path becomes significantly curved. This means that the slow-turn approximation we used to compute $n_s$ and $r$ is no longer valid, so the phenomenology of the models at the end of the trajectory will differ significantly in the two-field case vs. the single-field case.\footnote{Note that the slow-roll slow-turn approximation remains valid for all the models whose phenomenology is shown in figure \ref{fig:twofieldnsr}, so this figure is robust.} We leave an analysis of these models beyond the slow-roll slow-turn approximation for future study.


\section{Comparison with GRF Inflation \label{sec:COMPARISON}}

How do models of inflation learned via gradient ascent compare with those generated by other means? In this section, we compare our previous results to those of Gaussian random field (GRF) models \cite{Masoumi:2016eag, Bjorkmo:2017nzd, Bachlechner:2014rqa, Easther:2016ire}. In GRF models, the potential is taken to be a random field obeying Gaussian statistics:
\begin{align}
\langle V({\phi_1}) V({ \phi_2}) \rangle = \Lambda_v^8 e^{-|\phi_1-\phi_2|^2/{2 \Lambda_h^2}}:=  \frac{1}{(2 \pi)^N} \int d^N k P(k) e^{i {\bf k} \cdot (\phi_1 - \phi_2)},
\label{eq:GRF}
\end{align}
with
\begin{align}
P(k) =  \Lambda_v^8 (2 \pi \Lambda_h^2)^{N/2} e^{-\Lambda_h^2 k^2/2}.
\end{align}
with $\Lambda_h$ setting the correlation length, $\Lambda_v$ setting the height of the potential (which again drops out of inflationary observables $n_s$, $r$ in the slow-roll approximation), and $N$ the number of fields. We will focus on single-field GRF models, so we set $N=1$. Taylor expanding the potential as in (\ref{eq:pot}),
\begin{equation}
V(\phi) = \sum_{i=2}^N \Lambda_v^4 c_i \frac{\phi^i}{\Lambda_h^i},
\end{equation}
we can translate the distribution on $V(\phi)$ in (\ref{eq:GRF}) to a distribution on the coefficients $c_i$. For details on how this is done, we refer the reader to \cite{Masoumi:2016eag}, p. 3-7.

\begin{figure}[t!]
\begin{center}
\includegraphics[trim={0cm 0cm 0cm 0cm},clip,scale=0.6]{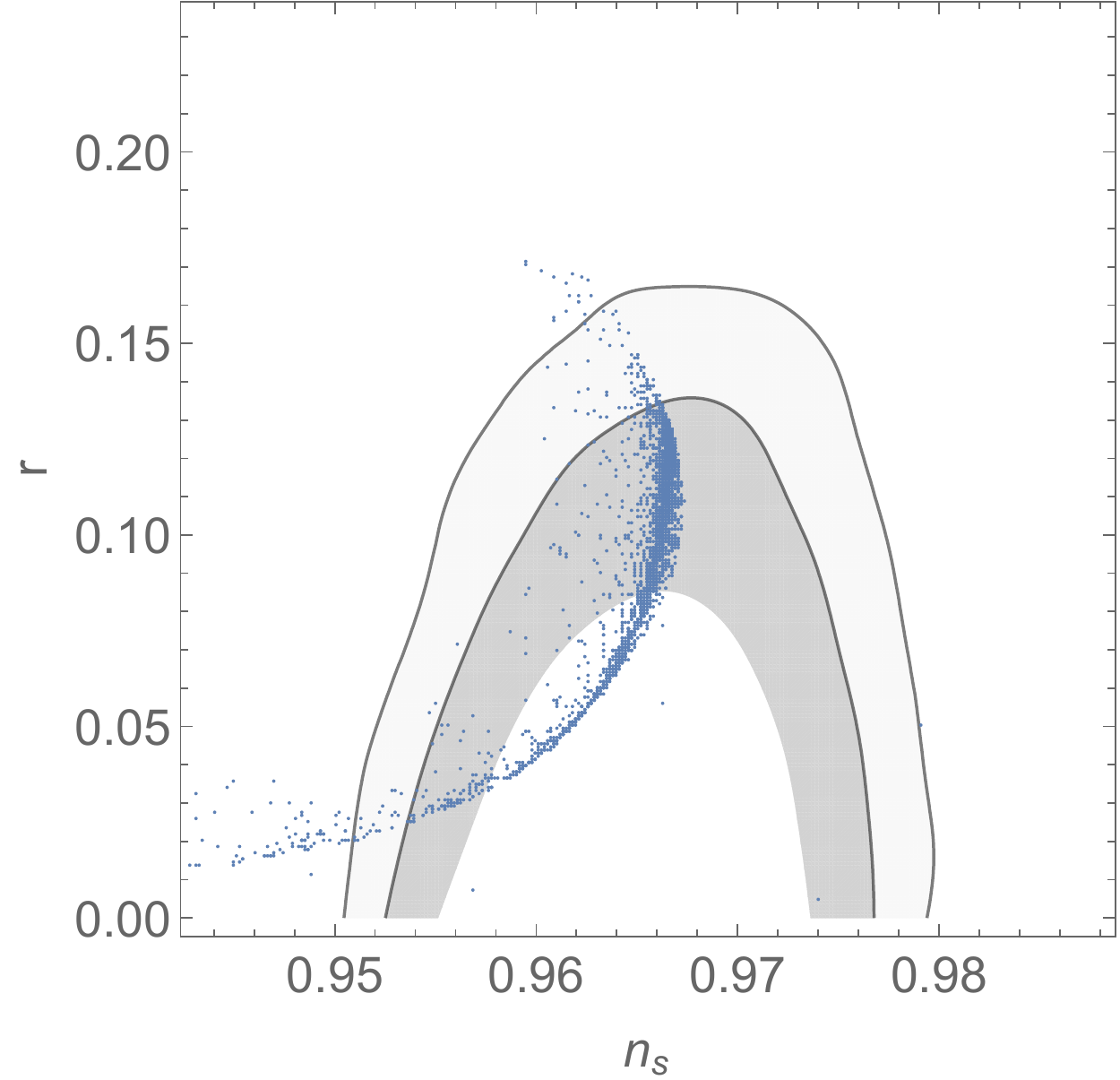}
\end{center}
\caption{Phenomenology of GRF models with $1 \leq \Lambda_h \leq 60$. The trajectory is qualitatively similar to that of gradient ascent models, though the variance is smaller.}
\label{fig:GRFnsr}
\end{figure}

To compare with our gradient ascent models, we set $N=13$ and generate $100$ potentials with $\Lambda_h$ taking all integer values between $1$ and $60$. For the models that produce at least 60 $e$-folds of inflation, we compute the values of $n_s$, $r$ and plot them in figure \ref{fig:GRFnsr}. We see the same qualitative features we have come to expect from random inflation models (decreasing $\Lambda_h$ leads to decreasing $r$, $n_s$), though the data points from the GRF models seem to be packed closer together than we saw in gradient ascent models.

\begin{figure}[t!]
\begin{center}
\begin{subfigure}[b]{0.45\textwidth}
\includegraphics[trim={0cm 0cm 0cm 0cm},clip,scale=0.5]{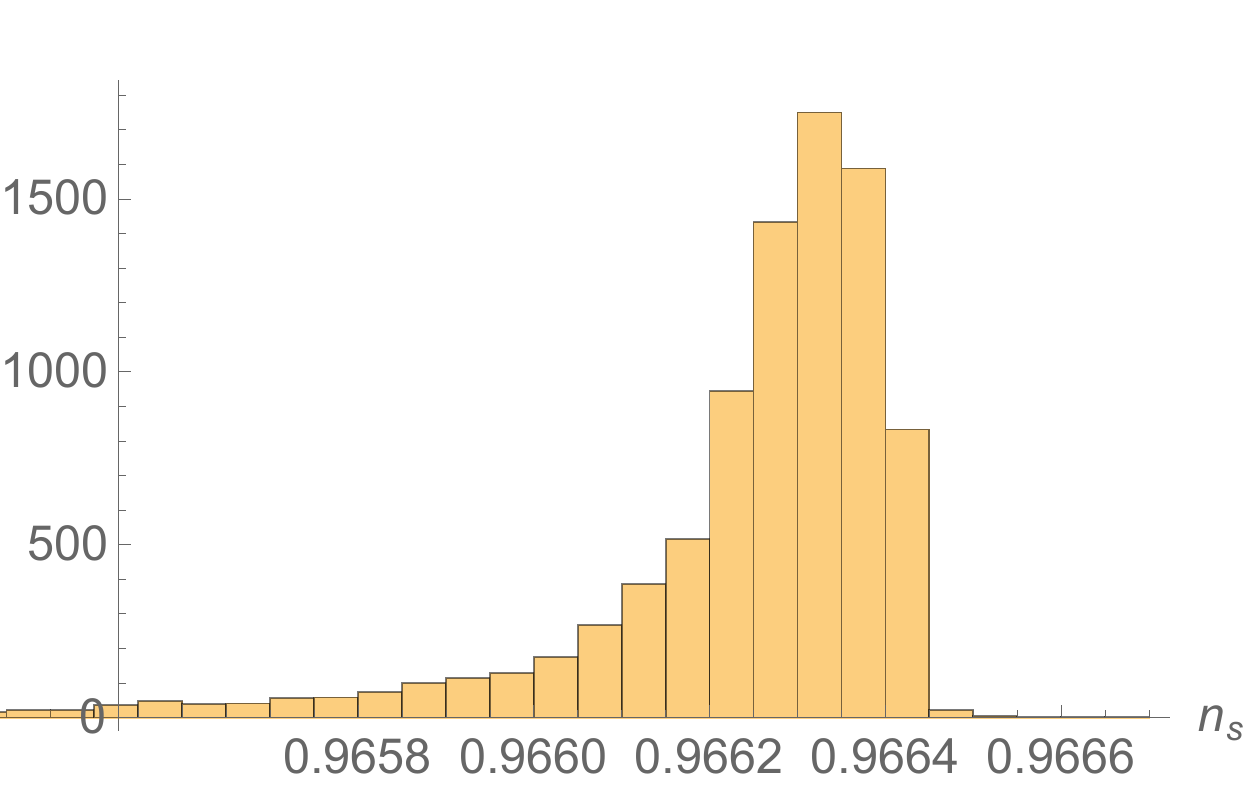}
\caption{GRF}
\end{subfigure}
\begin{subfigure}[b]{0.45\textwidth}
\includegraphics[trim={0cm 0cm 0cm 0cm},clip,scale=0.5]{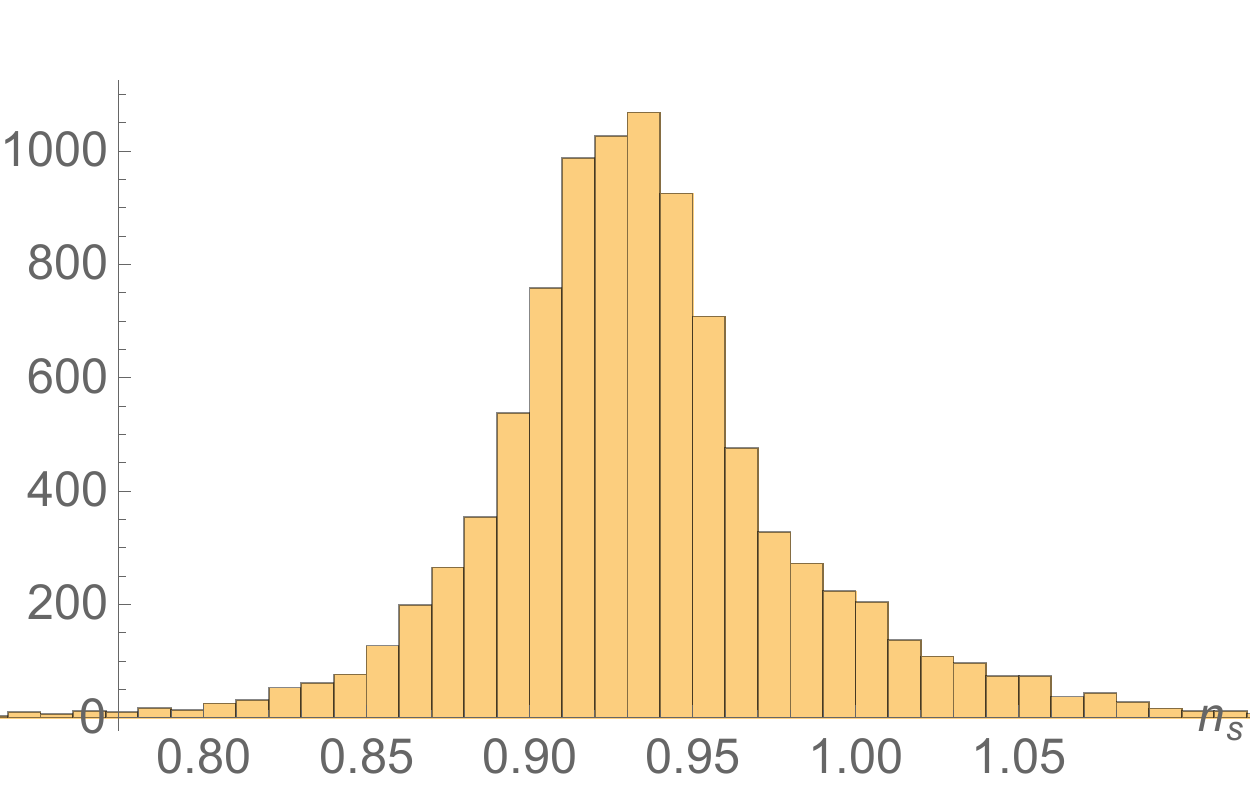}
\caption{Gradient ascent}
\end{subfigure}
\end{center}
\caption{Histogram of $n_s$ values for $\Lambda_h = 25 M_{\rm Pl}$ under the various approaches to random inflation. GRF models produce a much smaller variance than do either approaches to gradient ascent.}
\label{fig:GRFgaspread}
\end{figure}

We can make this precise by analyzing the gradient ascent trajectories at a fixed scale $\Lambda_h$, which we take to be $25 M_{\rm Pl}$. We do this as follows: for every gradient ascent trajectory, we choose the point whose $\Lambda_h$ value is closest to $25 M_{\rm Pl}$ (under approach 1), restricted to those with $ |\Lambda_h - 25 M_{\rm Pl} | < 0.1 M_{\rm Pl} $. We similarly generate a set of GRF models with $\Lambda_h = 25 M_{\rm Pl}$ and compare the $n_s$ values of the gradient ascent models with those of the GRF models.

Figure \ref{fig:GRFgaspread} shows the result of this comparison. If we ignore outlying models with $n_s < 0.5$ (at which point the second slow-roll parameter $\eta_V$ starts to grow large, and the slow-roll approximation in invalidated) the GRF approach yields a standard deviation of $\Delta n_s = 0.015$, whereas gradient ascent (approach 1) yields $\Delta n_s = 0.051$. Note also that the GRF distribution is skewed left, whereas the gradient ascent distribution is approximately symmetric.


\section{Conclusions \label{sec:CONC}}

We have examined the phenomenological predictions of a model that ``learns to inflate." Starting from random initial conditions, we computed a trajectory in the space of Wilsonian coefficients by performing a gradient ascent in the number of $e$-folds. We found that the corresponding inflationary potentials frequently yield phenomenology in agreement with experiment.

As with all approaches to random inflation, our approach is valuable only insofar as it captures universal features of inflation. Whether or not this is the case is a very difficult question given our rudimentary knowledge of inflation and the string landscape. However, one interesting point in this regard is the similarity between large-field inflationary models generated by gradient ascent in $N_e$, GRF models, and models of natural inflation \cite{Freese:1990rb}. Models of natural inflation are parametrized by an axion decay constant $f$, which controls the width of the cosine and correspondingly the distance traversed by the inflaton $\Delta \phi$. Natural inflation models with $f \gtrsim 30 M_{\rm Pl}$ yield a tensor-to-scalar ratio that is too large to agree with experiment, while models with $f \lesssim 15 M_{\rm Pl} $ yield a spectral index that is too small. This is quite reminiscent of our results in sections \ref{sec:RESULTS} and \ref{sec:COMPARISON}. Indeed, parametrized as a function of $f$, the trajectory of natural inflation in the $n_s$-$r$ plane looks almost identical to one of the trajectories in figures \ref{fig:trajectories} and \ref{fig:GRFnsr}, and the trajectories in figure \ref{fig:twofieldnsr} look similar but blueshifted. In this sense, one might say that our (large-field) gradient ascent models, GRF models, and natural inflation models lie in the same ``universality class." It is also worth noting that universal behavior makes individual models--such as natural inflation--harder to test; a precise measurement of $n_s = 0.96$, $r =0.05$ would be consistent with natural inflation, but it would also be consistent with our models produced by gradient ascent. And if gradient ascent is capable of producing said values with ease, one might suspect that other mechanisms--not just sinusoidal axion potentials produced by instantons--might also be able to produce it. Likewise, in the small-field case, our gradient ascent led to a family of inflection point models, similar to other approaches to random inflation \cite{McAllister:2012am}.

In the two-field case, we found that gradient ascent ultimately produced a genuine multi-field model of inflation, marked by significant curvature of the inflaton path in the vast majority of cases. This differs from other approaches to random multi-field inflation, such as that of \cite{Bjorkmo:2017nzd}, which typically required $\gtrsim 10$ fields to generate significant multi-field effects. We also saw that $N$-flation, which involves inflating along the diagonal direction of two identical fields, represents the stable direction of a saddle point in the $e$-folds function $\tilde N_e(\{ c_{ij} \})$.

There are many possible modifications of our approach, which could be promising areas of future study. Perhaps the most obvious and pressing one is to generalize from one- and two-field inflation to $N$-field inflation. String compactifications typically give $N \sim O(100)$, so it is important to understand what novel and universal behavior might arise in inflationary models in the large $N$ limit. Further, GRF and DBM models with $N \gg 1$ differ significantly from those with $N=1$, making it reasonable to think that large $N$ gradient ascent models will differ significantly from the ones considered here. The practical difficulty with such an analysis is that the number of terms in the inflationary potential grows exponentially with $N$, making such studies computationally expensive. However, it is precisely in such situations, in which the number of weights (i.e. Wilsonian coefficients) grows large, that the full power of modern machine learning may prove invaluable. In particular, we used simple gradient ascent to maximize the number of $e$-folds, but more complicated optimization methods could also be used. These methods could significantly decrease the time it takes to learn a model of inflation, especially in the many-field case.

\begin{figure}[t!]
\begin{center}
\includegraphics[trim={0cm 0cm 0cm 0cm},clip,scale=0.5]{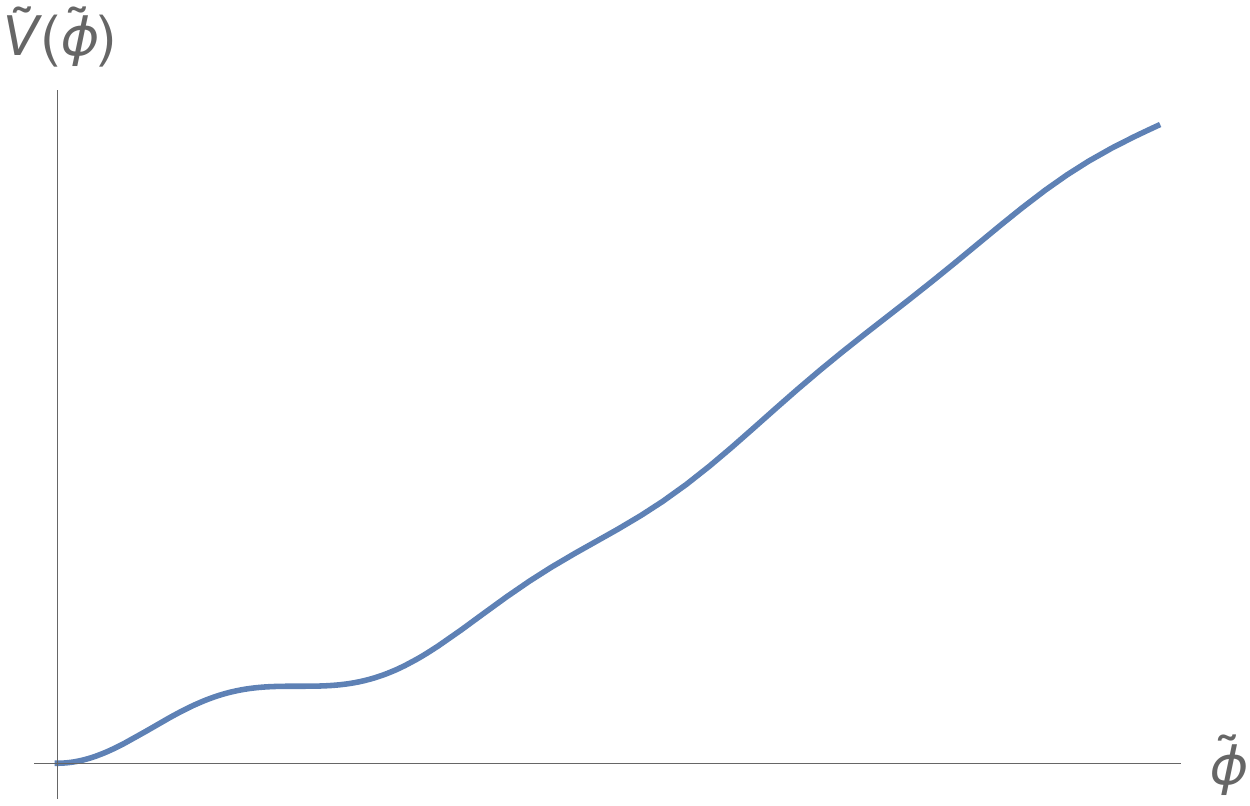}
\end{center}
\caption{Part of a periodic potential learned by gradient ascent in the space of Fourier mode coefficients. Note that this potential has multiple plateaus.}
\label{fig:periodic}
\end{figure}

We used a trivial kinetic term for the inflaton, but one could consider more general ones. It would be interesting to see if this approach might produce a model of DBI inflation \cite{Silverstein:2003hf} or something similar. Another possible modification is to consider an expansion in Fourier modes rather than monomials. Namely, one might consider a potential of the form
\begin{equation}
V(\phi) = \Lambda_v^4 \left( \sum_i a_i  \cos (i \phi/f) + b_i  \sin (i \phi/f)   \right)
\end{equation}
and perform a gradient ascent in the space of coefficients $\{ a_i, b_i \}$ to maximize the number of $e$-folds. Preliminary investigations of such models reveal that the potentials ``learned" in this way typically have multiple plateaus or other jagged features, as shown in figure \ref{fig:periodic}. These features could be an unwelcome indication that the model is wrong, but they might also produce interesting features in the CMB power spectrum that could be tested with experiment.

Alternative approaches to random inflation frequently encounter regions of parameter space (such as $\Lambda_h < M_{\rm Pl}$) where significant inflation is rare, making it computationally expensive to generate enough inflationary potentials to perform a statistical analysis. In these cases, combining said approach with gradient ascent on the number of $e$-folds could be useful for generating more inflationary potentials with $\gtrsim 50$ $e$-folds of inflation, reducing the computational cost.

Finally, we sought to maximize the number of $e$-folds, but one could consider alternative reward functions. For instance, taking a bottom-up approach, one might want to condition on the observed value of the spectral index and tensor-to-scalar ratio by assigning higher value to models that agree better with experiment i.e. incorporating the \emph{Planck} likelihood of the model into the reward function. From a top-down perspective, one might hope that improved understanding of the string landscape or anthropics could provide additional insight that may be incorporated into the reward function. More generally, it would interesting to see if other proposed measures for random inflation could be translated to reward functions for a model of gradient ascent.

Our approach represents just one application of machine learning techniques to theoretical cosmology. It is our hope and expectation that further applications will prove invaluable in addressing the wide array of difficult problems in cosmology and quantum gravity.

\section*{Acknowledgements}

We thank Liam McAllister and Matthew Reece for comments on an earlier draft of this article, and Kyle Cranmer, Henry Lin, and Eva Silverstein for helpful discussions. The author is supported by the Carl P. Feinberg Founders
Circle Membership and by NSF grant PHY-1606531.


\appendix

\section{Gradient Ascent }

The quintessential problem in machine learning is to minimize (maximize) a cost (reward) function, which is a map from a set of inputs or events into the real numbers. For instance, in a neural network, a single neuron consists of a map $f$ from some linear combination of inputs $x_1, x_2,...,x_k$ to some real number $y$:
\begin{equation}
y=f(w_1 x_1 + w_2 x_2 + ...+w_k x_k).
\end{equation}
Here, the coefficients $w_i$ are called weights. In a simple neural network consisting of this single neuron, the output $y$ may be compared against some observed value $y_{\text{obs}}$, and the ``cost function" $C$ may be defined as the square of the difference between the predicted and observed values, summed over all pairs of data $(\{x_i^{(j)} \} , y^{(j)})$:
\begin{equation}
C( \{ w_i\}) = \sum_j \Big(  f(w_1 x^{(j)}_1 + w_2 x^{(j)}_2 + ...+w_k x^{(j)}_k) -y^{(j)}_{\text{obs}}(\{ x^{(j)}_1,...,x^{(j)}_k\}) \Big)^2.
\label{eq:costfunction}
\end{equation}
The cost function effectively measures the predictive power of the model. For a particular set of inputs $\{x_i\}$ and output $y_{\text{obs}}$, the cost function is a function of the weights $\{w_i\}$. Training the neural network amounts to determining the weights $\{w_i\}$ that minimize the cost function.

In practice, this minimization procedure is typically carried out through the process of gradient descent or some related method. To begin, the weights $w_i$ are initialized to random values $w_i^{(0)}$. After this, subsequent values of the cost function are determined recursively according to the formula,
\begin{equation}
w_i^{(j+1)} = w_i^{(j)} - \eta\, \frac{\partial}{\partial w_i} C( \{ w_i^{(j)}  \} )   ~,~j \geq 0.
\end{equation}
Here, $\eta$ is a parameter known as the ``learning rate," which controls how drastically the parameters of the model are allowed to change at each step. The gradient $\partial_i C( \{ w_i^{(j)}  \} ) $ is estimated by summing over all data points $(\{x_i^{(j)} \} , y^{(j)})$ as in (\ref{eq:costfunction}).\footnote{Alternatively, in \emph{stochastic} gradient descent, the gradient $\partial_i C$ is estimated by summing over a random sample of data points. Stochastic gradient descent is much more efficient that ordinary gradient descent for large datasets.}

In the case of a reward function $R( \{ w_i\})$, the goal is instead to maximize with respect to the weights. In this case, we use a gradient ascent rather than a gradient descent,
\begin{equation}
w_i^{(j+1)} = w_i^{(j)} + \eta\, \frac{\partial}{\partial w_i} R( \{ w_i^{(j)}  \} )   ~,~j \geq 0.
\end{equation}

For an introduction to machine learning aimed at physicists, the reader may wish to consult reference \cite{Mehta:2018dln}.

\section[Single-Field Slow Roll Inflation]{Single-Field Slow Roll Inflation\\ {\large (This Slope Is Treacherous)}}

Cosmological inflation \cite{Guth:1980zm,Albrecht:1982wi,Linde:1981mu} is a proposed paradigm of the early universe in which quasi-exponential expansion is driven by the potential energy of one or more scalar fields. Concentrating on the case of a single scalar field, or ``inflaton," the dynamics are governed by the field equations
\begin{align}
\ddot{\phi} + 3 H \dot \phi + \frac{\partial V}{\partial \phi}&=0  \label{eq:motion} \\
 \frac{1}{3} \left( \frac{1}{2}\dot \phi^2 + V(\phi) \right)&=H^2. \label{eq:FRW}
\end{align}
Here, $V(\phi)$ is the potential of the field $\phi$, $H = \dot a /a$ measures the rate of expansion of the universe, and we have set the reduced Planck mass $M_{\rm Pl}$ equal to 1. Successful inflation requires $H$ to vary slowly with time, which implies that $\dot \phi$ is approximately constant, so $\ddot \phi$ can be neglected. As a result the first term in (\ref{eq:motion}) disappears, and the dynamics of $\phi$ effectively become a simple gradient descent towards a minimum of the potential:
\begin{equation}
\dot \phi =- \frac{1}{3 H} \frac{\partial V}{\partial \phi}.
\end{equation}
To quantify the validity of the ``slow-roll" approximation used above, we can define ``slow-roll parameters":
\begin{align}
\epsilon_H = - \frac{\dot H}{H^2}  ~,~~
\eta_H = - \frac{\ddot H}{2 H \dot H}  ~,~~
\xi_H = \frac{\dddot{H}}{2 H^2 \dot H} - 2 \eta_H^2 \label{eq:Hparam} 
\end{align}
Successful slow-roll inflation requires the slow-roll parameters to be much smaller than 1. When this holds, we can translate these slow-roll conditions to conditions on the potential, defining
\begin{align}
\epsilon_V =  \frac{1}{2} \frac{{V'}^2}{V^2}~,~~
\eta_V =  \frac{V''}{V} ~,~~\xi_V = \frac{V' V'''}{V^2}. \label{eq:Vparam} 
\end{align}
When these slow-roll parameters are small, they can be related by
\begin{equation}
\epsilon_V \approx \epsilon_H~,~~\eta_V \approx \eta_H + \epsilon_H~,~~\xi_V \approx \xi_H + 3 \epsilon_H \eta_H.
\end{equation}
Inflation ends when $\epsilon_V \approx 1$. The number of $e$-folds of inflation, defined by $N_e = \log a_{\rm end}/a_{\rm start}$, is given approximately by
\begin{equation}
N_e = \int_{\phi_{\rm end}}^{\phi_{\rm start}} \frac{d\phi}{\sqrt{2 \epsilon_V}}.
\end{equation}
Agreement with observation requires at least 50-60 $e$-folds of inflation. Measurable quantities like the tensor-to-scalar ratio $r$, the spectral index $n_s$, and the running of the spectral index $\alpha_s$ can be related to the slow-roll parameters via
\begin{equation}
r = 16 \epsilon_V^*~,~~n_s = 1+2 \eta_V^*-6 \epsilon_V^* ~,~~\alpha_s = -2 \xi_V^* + 16 \epsilon_V^* \eta_V^* -24 (\epsilon_V^*)^2.
\end{equation}
Here, the $^*$s indicate the values of the quantities 50-60 $e$-folds before the end of inflation. In this paper, we take the upper limit of 60 $e$-folds for the sake of concreteness.

More information on single-field slow-roll inflation can be found in \cite{Baumann:2009ds}.

\section[Two-Field Slow-Roll Inflation]{Two-Field Slow-Roll Inflation\\ {\large (This Path is Reckless)}}\label{sec:twofieldapp}

Two-field dynamics with a trivial field space metric are governed by the equations of motion
\begin{align}
\ddot{\phi}^i + 3 H \dot \phi^i + \frac{\partial V}{\partial \phi^i}&=0  \label{eq:motion2} \\
 \frac{1}{3} \left( \frac{1}{2}|\dot \phi|^2 + V(\phi^1, \phi^2) \right)&=H^2.
 \label{eq:twofieldFRW}
\end{align}
with $V(\phi^1,\phi^2)$ the two-field potential. In single-field inflation, we focused on the slow-roll regime, in which the inflaton rolls slowly down to the minimum of its potential. In two-field inflation, we work in the slow-roll, slow-turn regime, in which the inflaton rolls slowly and also turns slowly. In this approximation, we can neglect the acceleration term and approximate the dynamics with
\begin{equation}
\dot \phi^i = -\frac{1}{3H} \frac{\partial V}{\partial \phi^i}.
\end{equation}
Once again, the dynamics are effectively a gradient descent. We can again define slow-roll parameters. The first slow-roll parameter generalizes simply,
\begin{equation}
\epsilon_H = - \frac{\dot H}{H^2}.
\end{equation}
However, the second slow-roll parameter becomes a two-component vector in the two-field case. Following \cite{Peterson:2010np}, we define
\begin{equation}
\eta^i_H = \frac{\ddot\phi^i}{H^2}- \frac{\dot H \dot \phi^i}{H^3}.
\end{equation}
As in the single-field case, we can relate these to the shape of the potential. We define
\begin{equation}
\epsilon= \frac{1}{2}\left( \frac{|\partial_i V|^2}{V^2} \right).
\end{equation}
Here and henceforth, we drop the subscript $_V$ on the slow-roll parameters defined in terms of the potential. We also define
\begin{equation}
M_{ij} = \partial_i \partial_j  \log V.
\end{equation}
At first order in the slow-roll slow-turn approximation, we then have
\begin{equation}
\eta^i = \sum_{j} M_{ij} \partial_j (\log V).
\end{equation}
We then define the field speed $v = \sqrt{2 \epsilon_V}$, the unit vector $\bf{e}_{\parallel}$ in the direction of the gradient ${ \bf \nabla} V$, and a perpendicular unit vector $\bf{e}_\perp$. At first order, we can then write
\begin{align}
\frac{\eta_\parallel}{v} &= - \sum_{i,j} e_{\parallel}^i M_{ij} e_{\parallel}^j := - M_{\parallel \parallel}, \\
\frac{\eta_\perp}{v} &= - \sum_{i,j} e_{\parallel}^i M_{ij} e_{\perp}^j := - M_{\parallel \perp}.
\end{align}
$\eta_\parallel /v$ then measures the speed-up rate, while $\eta_\perp/v$ measures the turn rate.

As in the single-field case, we want to relate the inflationary potential $V$ to observable quantities. For this, we need to define a unit vector ${\bf e}_N$ in the direction of the gradient of the number of $e$-folds, ${\bf \nabla} N_e(\phi^1,\phi^2)$. We then define $\Delta_N$ by
\begin{equation}
\bfe_N = \cos \Delta_N \bfe_\parallel + \sin \Delta_N \bfe_\perp.
\end{equation} 
Then, at first order in slow-roll slow-turn, we can relate the slow-roll parameters to the tensor-to-scalar ratio $r$ and the spectral index $n_s$ by\footnote{$n_s$ here is the spectral index for the entropic power spectrum. In this paper, we ignore the power spectrum of isocurvature modes, as they are negligible in the models we consider. Generically, multi-field models of inflation can produce a significant spectrum of isocurvature modes, and these can mix with entropic modes.} \cite{Peterson:2010np}
\begin{align}
r &= 16 \epsilon \cos^2 \Delta_N \\
n_s &= 1 - 2 \epsilon +2 \sum_{i,j} e_N^i M_{ij} e_N^j.
\end{align}
Note that these quantities implicitly depend on the entire trajectory of the inflaton through the function $N_e$ and cannot be determined simply through a local analysis of the potential.

We can also write an expression for the parameter $f_{\text{NL}}^{\text{local}}$, which measures the non-Gaussianity of the scalar power spectrum \cite{Peterson:2010mv}:
\begin{align}
- \frac{6}{5} f_{\text{NL}}^{\text{local}} = \frac{\sum_{i,j}(\partial_i N_e) (\partial_i \partial_j N_e) (\partial_j N_e)}{|\partial_i N_e|^4}.
\end{align}


\newpage

\bibliographystyle{utphys}
\bibliography{ref}

\end{document}